# Sub-Nyquist Sampling: Bridging Theory and Practice[1]

Moshe Mishali and Yonina C. Eldar

[ A review of past and recent strategies for sub-Nyquist sampling ]

Signal processing methods have changed substantially over the last several decades. In modern applications, an increasing number of functions is being pushed forward to sophisticated software algorithms, leaving only delicate finely-tuned tasks for the circuit level. Sampling theory, the gate to the digital world, is the key enabling this revolution, encompassing all aspects related to the conversion of continuous-time signals to discrete streams of numbers. The famous Shannon-Nyquist theorem has become a landmark: a mathematical statement which has had one of the most profound impacts on industrial development of digital signal processing (DSP) systems.

Over the years, theory and practice in the field of sampling have developed in parallel routes. Contributions by many research groups suggest a multitude of methods, other than uniform sampling, to acquire analog signals [1]–[6]. The math has deepened, leading to abstract signal spaces and innovative sampling techniques. Within generalized sampling theory, bandlimited signals have no special preference, other than historic. At the same time, the market adhered to the Nyquist paradigm; state-of-the-art analog to digital conversion (ADC) devices provide values of their input at equalispaced time points [7], [8]. The footprints of Shannon-Nyquist are evident whenever conversion to digital takes place in commercial applications.

Today, seven decades after Shannon published his landmark result in [9], we are witnessing the outset of an interesting trend. Advances in related fields, such as wideband communication and radio-frequency (RF) technology, open a considerable gap with ADC devices. Conversion speeds which are twice the signal's maximal frequency component have become more and more difficult to obtain. Consequently, alternatives to high rate sampling are drawing considerable attention in both academia and industry.

This work has been submitted to the IEEE for possible publication. Copyright may be transferred without notice, after which this version may no longer be accessible.

The authors are with the Technion—Israel Institute of Technology, Haifa 32000, Israel. Emails: moshiko@tx.technion.ac.il, yonina@ee.technion.ac.il.

M. Mishali is supported by the Adams Fellowship Program of the Israel Academy of Sciences and Humanities. Y. C. Eldar is currently on leave at Stanford, USA. Her work was supported in part by the Israel Science Foundation under Grant no. 170/10 and by the European Commission in the framework of the FP7 Network of Excellence in Wireless COMmunications NEWCOM++ (contract no. 216715).

In this paper, we review sampling strategies which target reduction of the ADC rate below Nyquist. Our survey covers classic works from the early 50's of the previous century through recent publications from the past several years. The prime focus is bridging theory and practice, that is to pinpoint the potential of sub-Nyquist strategies to emerge from the math to the hardware. In this spirit, we integrate contemporary theoretical viewpoints, which study signal modeling in a union of subspaces, together with a taste of practical aspects, namely how the avant-garde modalities boil down to concrete signal processing systems. Our hope is that this presentation style will attract the interest of both researchers and engineers with the aim of promoting the sub-Nyquist premise into practical applications, and encouraging further research into this exciting new frontier.

─── *Introduction* ───

We live in a digital world. Tele-communication, entertainment, gadgets, business – all revolve around digital devices. These miniature sophisticated black-boxes process streams of bits accurately at high speeds. Nowadays, electronic consumers feel natural that a media player shows their favorite movie, or that their surround system synthesizes pure acoustics, as if sitting in the orchestra, and not in the living room. The digital world plays a fundamental role in our everyday routine, to such a point that we almost forget that we cannot "hear" or "watch" these streams of bits, running behind the scenes. The world around us is analog, yet almost all modern man-made means for exchanging information are digital. "I am an analog girl in a digital world", sings Judi Gorman [One Sky, 1998], capturing the essence of the digital revolution.

ADC technology lies at the heart of this revolution. ADC devices translate physical information into a stream of numbers, enabling digital processing by sophisticated software algorithms. The ADC task is inherently intricate: its hardware must hold a snapshot of a fast-varying input signal steady, while acquiring measurements. Since these measurements are spaced in time, the values between consecutive snapshots are lost. In general, therefore, there is no way to recover the analog input unless some prior on its structure is incorporated.

A common approach in engineering is to assume that the signal is bandlimited, meaning that the spectral contents are confined to a maximal frequency $f_{\max}$. Bandlimited signals have limited (hence slow) time variation, and can therefore be perfectly reconstructed from equalispaced samples with rate at least $2f_{\max}$, termed the Nyquist rate. This fundamental result is often attributed in the engineering community to Shannon-Nyquist [9], [10], although it dates back to earlier works by Whittaker [11] and Kotelńikov [12].

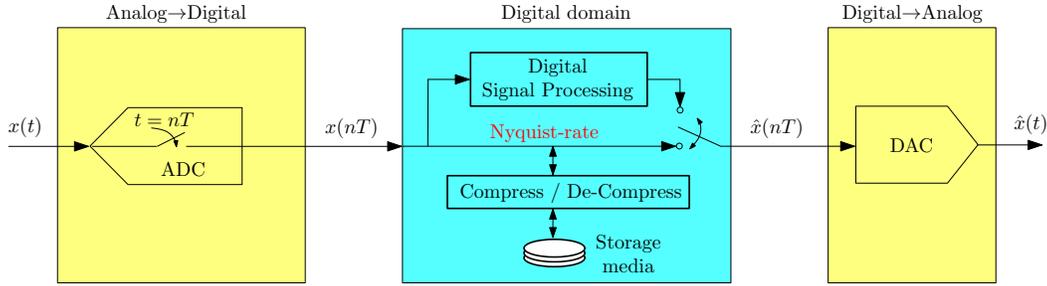

Fig. 1: Conventional blocks in a DSP system.

In a typical signal processing system, a Nyquist ADC device provides uniformly-spaced pointwise samples $x(nT)$ of the analog input $x(t)$, as depicted in Fig. 1. In the digital domain, the stream of numbers is either processed or stored. Compression is often used to reduce storage volume. DSP, which is unquestionably the crowning glory of this flow, is typically performed on the uncompressed stream. The delicate interaction with the continuous world is isolated to the ADC stage, so that sophisticated algorithms can be developed in a flexible software environment. The flow of Fig. 1 ends with a digital to analog (DAC) device which reconstructs $x(t)$ from the high Nyquist-rate sequence $x(nT)$.

A fundamental reason for processing at the Nyquist rate is the clear relation between the spectrum of $x(t)$ and that of $x(nT)$, so that digital operations can be easily substituted for their continuous counterparts. Digital filtering is an example where this relation is successfully exploited. Since the power spectral densities of continuous and discrete random processes are associated in a similar manner, estimation and detection of parameters of analog signals can be performed by DSP. In contrast, compression, in general, results in a nonlinear complicated relationship between $x(t)$ and the stored data.

This paper reviews alternatives to the scheme of Fig. 1, whose common denominator is sampling at a rate below Nyquist. Research on sub-Nyquist sampling spans several decades, and has been attracting renewed attention lately, since the growing interest in sampling in union of subspaces, finite rate of innovation (FRI) models and compressed sensing techniques. Our goal in this survey is to provide an overview of various sub-Nyquist approaches. We focus this presentation on one-dimensional signals $x(t)$, with applications to wideband communication, channel identification and spectrum analysis. Two-dimensional imaging applications are also briefly discussed.

Throughout, the theme is bridging theory and practice. Therefore, before detailing the specifics of various sub-Nyquist approaches, we first discuss the relation between theory and practice in a broader context. The example of uniform sampling, which without a doubt crossed that bridge, is used to list the essential ingredients of a sampling strategy so that it has the potential to step from math to actual hardware. Our subsequent presentation of sub-Nyquist strategies attempts to give a taste from both worlds – presenting the theoretical principles underlying each strategy and how they boil down to concrete and practical schemes. Where relevant, we shortly elaborate on practical considerations, *e.g.,* hardware complexity and computational aspects.

———— *Essential Ingredients of a Sampling System* ————

**Nyquist sampling**

In 1949, Shannon formulated the following theorem for "*a common knowledge in the communication art*" [9, Th. 1]:

> If a function $f(t)$ contains no frequencies higher than $W$ cycles-per-second, it is completely determined by giving its ordinates at a series of points spaced $1/2W$ seconds apart.

It is instructive to break this one-sentence formulation into three pieces. The theorem begins by defining an analog signal model – those functions $f(t)$ that do not contain frequencies above $W$ Hz. Then, it describes the sampling stage, namely pointwise equalispaced samples. In between, and to some extent implicitly, the required rate for this strategy is stated: at least $2W$ samples per second.

The bandlimited signal model is a natural choice to describe physical properties that are encountered in many applications. For example, a physical communication medium often dictates the maximal frequency that can be reliably transferred. Thus, material, length, dielectric properties, shielding and other electrical parameters define the maximal frequency $W$. Often, bandlimitedness is enforced by a lowpass filter with cutoff $W$, whose purpose is to reject thermal noise beyond frequencies of interest.

The implementation suggested by the Shannon-Nyquist theorem, equalispaced pointwise samples of the input, is essentially what industry has been persistently striving to achieve in ADC design. The sampling stage, per se, is insufficient; The digital stream of numbers needs to be tied together with a reconstruction algorithm. The famous interpolation formula

$$f(t) = \sum_n f\left(\frac{n}{2W}\right) \text{sinc}(2Wt - n), \qquad \text{sinc}(\alpha) \triangleq \frac{\sin(\pi\alpha)}{\pi\alpha}, \tag{1}$$

which is described in the proof of [9], completes the picture by providing a concrete reconstruction method. Although (1) theoretically requires infinitely many samples to recover $f(t)$ exactly, in practice, truncating the series to a finite number of terms reproduces $f(t)$ quite accurately [13].

| [Table 1] SUB-NYQUIST SAMPLING: A WISHLIST. | |
|---|---|
| **Ingredient** | **Requirement** |
| Signal model | Encountered in applications |
| Sampling rate | Approach the minimal for the model at hand |
| Implementation | Hardware: Low cost, small number of devices<br>Software: light computational loads, fast runtime |
| Robustness | React gracefully to design imperfections<br>Low sensitivity to noise |
| Processing | Allow various DSP tasks |

The theory ensures perfect reconstruction from samples at rate $2W$. A generalized sampling theorem by Papoulis allows to relax design constraints by replacing a single Nyquist-rate ADC by a filter-bank of $M$ branches, each sampled at rate $2W/M$ [14]. Another route to design simplification is oversampling, which is often used to replace the ideal brickwall filter by more flexible filter designs and to combat noise. Certain ADC designs, such as sigma-delta conversion, intentionally oversample the input signal, effectively trading sampling rate for higher quantization precision. Our wishlist, therefore, includes a similar guideline for sub-Nyquist strategies: achieve the lowest rate possible in an ideal noiseless setting, and relax design constraints by oversampling and parallel architectures.

Further to what is stated in the theorem, we believe that two additional ingredients motivate the widespread use of the Shannon theorem. First, the interpolation formula (1) is robust to various noise sources: quantization round-off, series truncation and jitter effects [13]. The second appeal of this theorem lies in the ability to shift processing tasks from the analog to the digital domain. DSP is perhaps the major driving force which supports the wide popularity of Nyquist sampling. In sub-Nyquist sampling, the digital stream is, by definition, different from the Nyquist-rate sequence $x(nT)$. Therefore, the challenge of reducing sampling rate creates another obstacle – interfacing the samples with DSP algorithms that are traditionally designed to work with the high-rate sequence $x(nT)$, without necessitating interpolation of the Nyquist-rate samples. In other words, we would like to perform DSP at the low sampling rate as well.

Table 1 summarizes a wishlist for a sub-Nyquist system, based on those properties observed in the Shannon theorem. A sampling strategy satisfying most of these properties can, hopefully, find its way into practical applications.

**Architecture of a sub-Nyquist system**

A high-level architecture of a sub-Nyquist system is depicted in Fig. 2, following the spirit of the traditional block diagram of Fig. 1. The ADC task is carried out by some hardware mechanism, which outputs a sequence $y[n]$ of measurements at a low rate. Since the sub-Nyquist samples $y[n]$ are, by definition, different from the uniform Nyquist sequence $x(nT)$ of Fig. 1, a digital core may be needed to preprocess the raw data before DSP can take place. A prominent advantage over conventional Nyquist architectures is that the DSP operations are carried out at the low input rate. The digital core may also be needed to assist in reconstructing $x(t)$ from $y[n]$. Another advantage is that storage may not require a preceding compression stage; conceptually, the compression has already been performed by the sub-Nyquist sampling hardware.

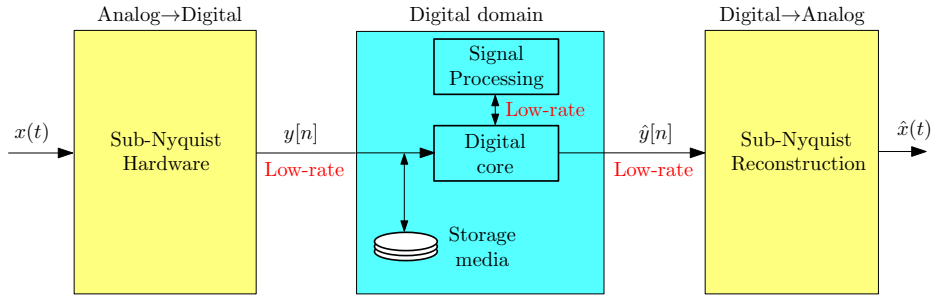

**Fig. 2:** A high-level architecture of a sub-Nyquist system. Both processing and continuous recovery are based on lowrate computations. The raw data can be directly stored.

An important point we would like to emphasize is that strictly speaking, none of the methods we survey actually breach the Shannon-Nyquist theorem. Sub-Nyquist techniques leverage known signal structure, that goes beyond knowledge of the maximal frequency component. The key to developing interesting sub-Nyquist strategies is to rely on structure that is not too limiting and still allows for a broad class of signals on the one hand, while enabling sampling rate reduction on the other. One of the earlier examples demonstrating how signal structure can lead to rate reduction is sampling of multiband signals with known center frequencies, namely, signals that consists of several known frequency bands. We begin our review with this classic setting. We then discuss more recent paradigms which enable sampling rate reduction even when the band positions are unknown. As we show, this setting is a special case of a more general class of signal structures known as unions of subspaces, which includes a variety of interesting examples. After introducing this general model, we consider several sub-Nyquist techniques which exploit such signal structure in sophisticated ways.

——— *Classic Sub-Nyquist Methods* ———

In this section we survey classic sampling techniques which reduce the sampling rate below Nyquist, assuming a multiband signal with known frequency support. An example of a multiband input with $N$ bands is depicted in Fig. 3, with individual bandwidths not greater than $B$ Hz, centered around carrier frequencies $f_i \leq f_{\max}$ ($N$ is even for real-valued inputs). Since the carriers $f_i$ are known and the spectral support is fixed, the set of multiband inputs on that support is closed under linear combinations, thereby forming a subspace of possible inputs. Overlapping bands are permitted, though in practical scenarios, *e.g.,* communication signals, the bands typically do not overlap.

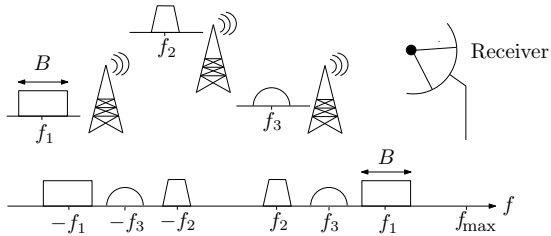
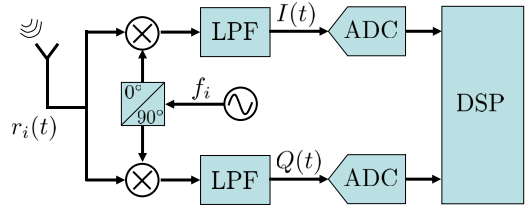

**Fig. 3:** Three RF transmissions with different carriers $f_i$. The receiver sees a multiband signal (bottom drawing).

**Fig. 4:** A block diagram of a typical I-Q demodulator.

**Demodulation**

The most common practice to avoid sampling at the Nyquist rate,

$$f_{\text{NYQ}} = 2f_{\text{max}}, \tag{2}$$

is demodulation. The signal $x(t)$ is multiplied by the carrier frequency $f_i$ of a band of interest, so as to shift contents of a single narrowband transmission from high frequencies to the origin. This multiplication also creates a narrowband image around $2f_i$. Lowpass filtering is used to retain only the baseband version, which is subsequently sampled uniformly in time. This procedure is carried out for each band individually.

Demodulation provides the DSP block with the information encoded in a band of interest. To make this statement more precise, we recall how modern communication is often performed. Two $B/2$-bandlimited information signals $I(t), Q(t)$ are modulated on a carrier frequency $f_i$ with a relative phase shift of $90°$. The quadrature output signal is then given by [15]

$$r_i(t) = I(t)\cos(2\pi f_i t) + Q(t)\sin(2\pi f_i t). \tag{3}$$

For example, in amplitude modulation (AM), the information of interest is the amplitude of $I(t)$, while $Q(t) = 0$. Phase- and frequency-modulation (PM/FM) obey (3) such that the analog message is $g(t) = \arctan(I(t)/Q(t))$ [16]. In digital communication, *e.g.,* phase- or frequency shift-keying (PSK/FSK), $I(t), Q(t)$ carry symbols. Each symbol encodes one, two or more 0/1 bits. The I/Q-demodulator, depicted in Fig. 4, basically reverts the actions performed at the transmitter side which constructed $r_i(t)$. Once $I(t), Q(t)$ are obtained by the hardware, a pair of lowrate ADC devices acquire uniform samples at rate $B$. The subsequent DSP block can infer the analog message or decode the bits from the received symbols.

Reconstruction of each $r_i(t)$, and consequently recovery of the multiband input $x(t)$, is as simple as remodulating the information on their carrier frequencies $f_i$, according to (3). This option is used in relay stations or regenerative repeaters which decode the information $I(t), Q(t)$, use digital error correction algorithms, and then transform the signal back to high frequencies for the next transmission section [17].

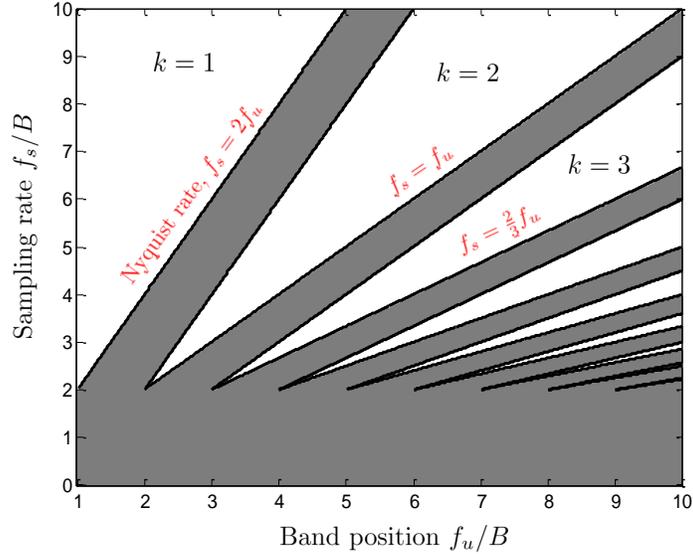

**Fig. 5:** The allowed (white) and forbidden (gray) undersampling rates of a bandpass signal depend on its spectral position [18].

I/Q demodulation has different names in the literature: zero-IF receiver, direct conversion, or homodyne; cf. [15] for various demodulation topologies. Each band of interest requires 2 hardware channels to extract the relevant $I(t), Q(t)$ signals. A similar principle is used in low-IF receivers, which demodulate a band of interest to low frequencies but not around the origin. Low-IF receivers require only one hardware channel per band, though the sampling rate is higher compared to zero-IF receivers.

**Undersampling ADC**

Aliasing is often considered an undesired effect of sampling. Indeed, when a bandlimited signal is sampled below its Nyquist rate, aliases of high-frequency content trample information located around other spectral locations and destroy the ability to recover the input. Undersampling (*a.k.a.*, direct bandpass sampling) refers to uniform sampling of a bandpass signal at a rate lower than the maximal frequency, in which case proper sampling rate selection renders aliasing advantageous.

Consider a bandpass input $x(t)$ whose information band lies in the frequency range $(f_l, f_u)$ of length $B = f_u - f_l$. In this case, the lowest rate possible is $2B$ [19]. Uniform sampling of $x(t)$ at a rate of $f_s$ that obeys

$$\frac{2f_u}{k} \leq f_s \leq \frac{2f_l}{k-1}, \tag{4}$$

for some integer $1 \leq k \leq f_u/B$, ensures that aliases of the positive and negative contents do not overlap [18]. Fig. 5 illustrates the valid sampling rates implied by (4). In particular, the figure and (4) show that $f_s = 2B$ is achieved only if $x(t)$ has an integer band positioning, $f_u = kB$. Furthermore, as the rate reduction factor $k$ increases, the valid region of sampling rates becomes narrower. For a given band position $f_u$, the region corresponding to the maximal $k \leq f_u/B$ is the most sensitive to slight deviations in the exact values of $f_s, f_l, f_u$ [18]. Consequently, besides the fact that $f_s = 2B$ cannot be achieved in general (even in ideal noiseless settings), a significantly higher rate is likely to be required in practice in order to cope with design imperfections.

Bridging theory and practice, the fact that (4) allows rate reduction, even though higher than the minimal, is useful in many applications. In undersampling, the ADC is applied directly to $x(t)$ with no preceding analog preprocessing components, in contrary to the RF hardware used in I/Q demodulation. However, not every ADC device fits an undersampling system: only those devices whose front-end analog bandwidth exceeds $f_u$ are viable. Box 1 expands on this constraint of front-end bandwidth in Nyquist and undersampling ADCs.

### Box 1. Nyquist and Undersampling ADC Devices

An ADC device, in the most basic form, repeatedly alternates between two states: track-and-hold (T/H) and quantization. During T/H, the ADC tracks the signal variation. When an accurate track is accomplished, the ADC holds the value steady so that the quantizer can convert the value into a finite representation. Both operations must end before the next signal value is acquired.

In the signal processing community, an ADC is often modeled as an ideal pointwise sampler that captures values of $x(t)$ at a constant rate of $r$ samples per second. As with any analog circuitry, the T/H function is limited in the range of frequencies it can accept: a lowpass filter with cutoff $b$ can be used to model the T/H capability [20].

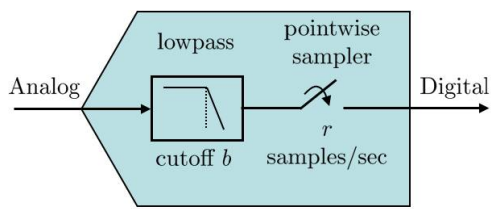

| Device | Max. rate (MSamples/sec) | Analog BW (MHz) |
|---|---|---|
| ADS5474 | 400 | 1440 |
| AD12401 | 400 | 480 |
| AD1020 | 105 | 250 |
| AD9433 | 105 | 750 |

In most off-the-shelf ADCs, the analog bandwidth parameter $b$ is specified higher than the maximal sampling rate $r$ of the device. The table lists example devices. When using an ADC at the Nyquist rate of the input, the filter can be omitted from the model, since the signal is

> bandlimited to $f_{\max} = r/2 \leq b$. In contrast, for sub-Nyquist purposes, the analog bandwidth $b$ becomes an important factor in accurate modeling and actual selection of the ADC, since it defines the maximal input frequency that can be undersampled:
>
> $$f_{\max} \leq b. \tag{5}$$
>
> Typically, $b$ specifies the $-3$ dB point of the T/H frequency response. Thus, if flat response in the passband is of interest, $f_{\max}$ cannot approach too close to $b$. For example, if $x(t)$ is a bandpass signal in the range $[600, 625]$ MHz, then undersampling at rate $f_s = 50$ MHz satisfies condition (4). In this example, whilst both AD9433 and AD10200 are capable of sampling at a rate $r \geq 50$ MHz, only the former is applicable due to (5).
>
> Undersampling ADCs have a wider spacing between consecutive samples. This advantage is translated into simplifying design constraints, especially in the duration allowed for quantization. However, regardless of the sampling rate $r$, the T/H stage must still hold a pointwise value of a fast-varying signal. In terms of analog bandwidth there is no substantial difference between Nyquist and undersampling ADCs; both have to accommodate the Nyquist rate of the input.

Undersampling has two prominent drawbacks. First, the resulting rate reduction is generally significantly higher than the minimal as evident from Fig. 5. As listed in Table 1, approaching the minimal rate, at least theoretically, is a desired property. Second, and more importantly, undersampling is not suited to multiband inputs. In this scenario, each individual band defines a range of valid values for $f_s$ according to (4). The sampling rate must be chosen in the intersection of these conditions. Moreover, it should also be verified that the aliases due to the different bands do not interfere. As noted in [21], satisfying all these constraints simultaneously, if possible, is likely to require a considerable rate increase.

**Periodic nonuniform sampling**

The discussion above suggests that uniform sampling may not be the most desirable acquisition strategy for inputs with multiband structure, unless sufficient analog hardware is used as in Fig. 4. Classic studies in sampling theory have focused on nonuniform alternatives. In 1967, Landau proved a lower bound on the sampling rate required for spectrally-sparse signals [19] with known frequency support when using pointwise sampling. In particular, Landau's theorem supports the intuitive expectation that a multiband signal $x(t)$ with $N$ information bands of individual widths $B$ necessitates a sampling rate no lower than the sum of the band widths, *i.e.*, $NB$.

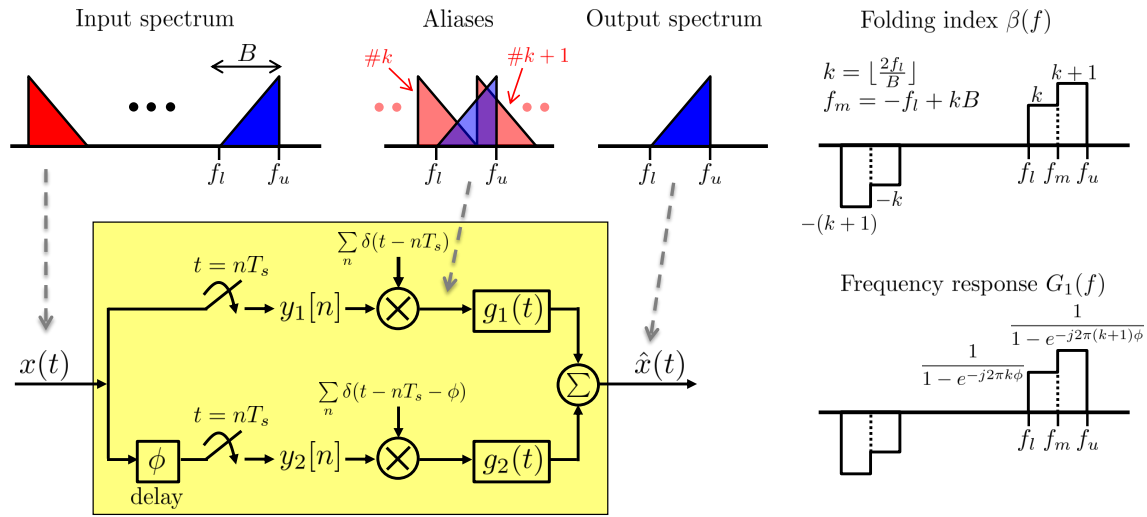

**Fig. 6:** Second-order PNS. The bandpass signal $x(t)$ is sampled by two rate-$B$ uniform sequences with relative time delay $\phi$. The interpolation filters cancel out the contribution of the undesired alias.

Periodic nonuniform sampling (PNS) allows to approach the minimal rate $NB$ without complicated analog preprocessing. Besides ADC devices, the hardware needs only a set of time-delay elements. PNS consists of $m$ undersampling sequences with relative time-shifts:

$$y_i[n] = x(nT_s + \phi_i), \quad 1 \le i \le m, \tag{6}$$

such that the total sampling rate $m/T_s$ is lower than $f_{\text{NYQ}}$. Kohlenberg [22] was the first to prove perfect recovery of a bandpass signal from PNS samples taken at an average rate of $2B$ samples/sec. Lin and Vaidyanathan [23] extended his approach to multiband signals.

We follow the presentation in [23] and explain how the parameters $m, T_s, \phi_i$ are chosen in the simpler case of a bandpass input. Suppose $x(t)$ is supported on $\mathcal{I} = (f_l, f_u) \cup (-f_u, -f_l)$ and $B = f_u - f_l$. We choose a PNS system with $m = 2$ channels (*a.k.a.,* second-order PNS), a sampling interval $T_s = 1/B$, $\phi_1 = 0$ and $\phi_2 = \phi$. Due to the undersampling in each channel, aliases of the band contents tile the spectrum, so that the positive and negative images fold on each other, as visualized in Fig. 6. In the frequency domain, the sample sequences (6) satisfy a linear system [23]

$$T_s Y_1(f) = X(f) + X(f - \beta(f)B), \tag{7a}$$

$$T_s Y_2(f) = X(f) + X(f - \beta(f)B)e^{-j2\pi\beta(f)\phi B}, \tag{7b}$$

for $f \in \mathcal{I}$. The function $\beta(f) = -\beta(-f)$ is piecewise constant over $f \in \mathcal{I}$, indexing the aliased images. The exact levels and transitions of $\beta(f)$ depend explicitly on the band position as shown in Fig. 6.

The aliases have unity weights in $y_1[n]$, whereas the time delay $\phi$ in $y_2[n]$ results in unequal weighting. System (7) is linearly independent as long as $\phi$ obeys

$$e^{-j2\pi\beta(f)\phi B} \neq 1. \tag{8}$$

Since $\beta(f)$ can take on only 4 distinct values within $f \in (f_l, f_u)$, there are many possible selections for $\phi$ which satisfy (8). Recovery of $x(t)$ is carried out by interpolation [22], [23]

$$x(t) = \sum_{n \in \mathbb{Z}} y_1[n]g_1(t - nT_s) + y_2[n]g_2(t - nT_s), \tag{9}$$

with bandpass filters $g_1(t), g_2(t)$, which reverse the weights in (7). These filters have frequency responses

$$G_1(f) = \frac{1}{1 - e^{-j2\pi\beta(f)\phi B}}, \quad G_2(f) = -G_1(f), \quad f \in \mathcal{I}, \tag{10}$$

as are drawn in Fig. 6. In practice, these filters can be realized digitally, so that the output of Fig. 6 is the Nyquist-rate sequence $x(nT)$, with $T = 1/2f_u$ equal to the Nyquist interval. Subsequently, a DAC device may interpolate the continuous signal $x(t)$.

The extension to multiband signals with $N$ bands of individual widths $B$ is accomplished following the same procedure using an $N$th order PNS system, with delays $\phi_l$, $1 \leq l \leq N$ [23]. Reconstruction consists of $N$ filters, which are piecewise constant over the frequency support of $x(t)$. The indexing function $\beta(f)$ is extended to an $N \times N$ matrix $\mathbf{A}(f)$, with entries depending on $\phi_l$ and band locations. In general, an $N$th-order PNS can resolve up to $N$ aliases, since it provides a set of $N$ equations. The equations are linearly independent, or solvable, if $\mathbf{A}^{-1}(f)$ exists over the entire multiband support [23]. Lin and Vaidyanathan show that the choice $\phi_l = l\phi$ renders $\mathbf{A}(f)$ a Vandermonde matrix, in which case the choice of the single delay $\phi$ is tractable. Bands of different widths are treated by viewing the bands as consisting of narrower intervals which are integer multiplies of a common length. For example, if $N = 4$ (two transmissions) and $B_1 = k_1 B, B_2 = k_2 B$, then the equivalent model has $4(k_1 + k_2)$ bands of equal width $B$. This conceptual step allows to achieve the Landau rate. For technical completeness, the same solution applies to mixed rational-irrational bandwidths for an infinitesimal rate increase.

**PNS vs. demodulation**

An apparent advantage of PNS over RF demodulation is that it can approach Landau's rate with no hardware components preceding the ADC device. This theoretical advantage, however, was not widely embraced by industry for acquisition of multiband inputs. In an attempt to reason this situation, we leverage practical insights from time-interleaving ADCs, a popular design topology

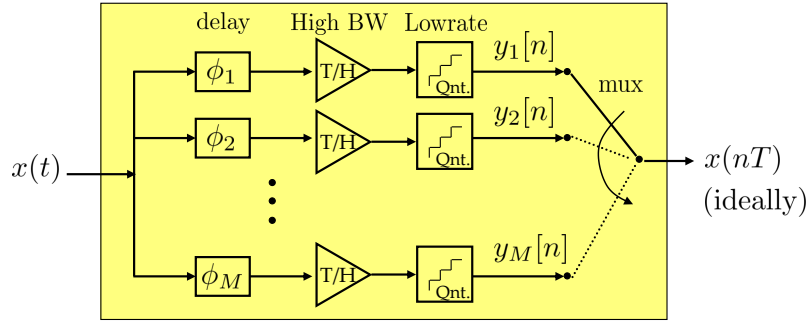

**Fig. 7:** Block diagram of a time-interleaved ADC.

used in high-speed converters [24]–[26].

Time-interleaved ADC technology splits the task of converting a wideband signal into $M$ parallel branches, essentially utilizing Papoulis' theorem with a bank of time-delay elements. Each branch in the block diagram of Fig. 7 introduces a time delay of $\phi_l$ seconds and subsequently samples $x(t - \phi_l)$ at rate $1/MT$, where $T = 1/f_{\text{NYQ}}$ is the Nyquist interval. Ideally, when $\phi_l = lT$, interleaving the $M$ digital streams provides a sequence that coincides with the Nyquist rate samples $x(nT)$. A time-interleaving ADC consists of $M$ separate T/H circuitries and quantizers, thereby relaxing design constraints by allowing each branch to perform the conversion task in a duration of $MT$ seconds rather than $T$. Whilst the larger duration simplifies quantization, the T/H complexity remains almost the same – it still needs to track a Nyquist-rate varying input and hold its value at a certain time point, regardless of the higher duration allocated for conversion, as explained in Box 1.

PNS is a degenerated time-interleaved ADC with only $m < M$ branches. This means that a PNS-based sub-Nyquist solution requires Nyquist-rate T/H circuitries, one per sampling branch. In addition to high analog bandwidth, PNS also requires compensating for imperfect production of the time delay elements. Consequently, realizing PNS in practice may not be much easier than designing an $M$-channel time-interleaved ADC with Nyquist-rate sampling capabilities. Thus, while time-interleaving is a popular design method for Nyquist ADCs, it may be less useful for the purpose of sub-Nyquist sampling of wideband signals with large $f_{\text{NYQ}}$.

More broadly, any pointwise strategy, which is applied directly on a wideband signal, has a technological barrier around the maximal rate of commercial T/H circuitry. This barrier creates an (undesired) coupling between advances in RF and ADC technologies; as transmission frequencies grow higher, a comparable speed-up of T/H bandwidth is required. With accelerated development of RF devices, a considerable gap has already been opened, rendering ADCs a bottleneck in many modern signal processing applications. In contrast, in demodulation, even though the signal is wideband, an ADC with low analog bandwidth is sufficient due to the preceding lowpass filter. RF preprocessing (mixers and filters) buffer between $x(t)$ and actual ADCs, thereby offering a scalable sampling solution, which effectively decouples T/H capabilities from dependency on the input's maximal frequency. More importantly, demodulation ensures that only in-band noise enters the system, whereas in PNS, out-of-band noise from the entire Nyquist bandwidth is aggregated.

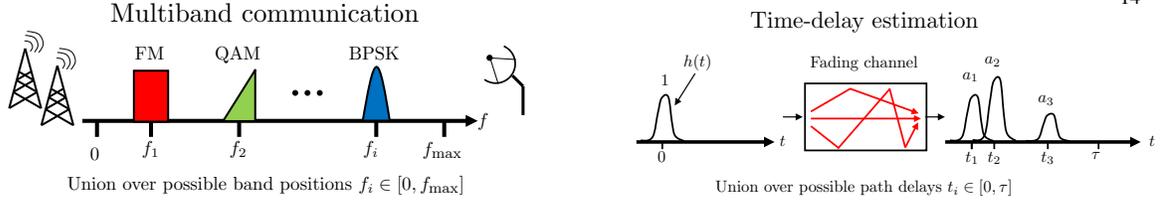

**Fig. 8:** Example applications of UoS modeling.

We now turn to review sub-Nyquist techniques when the carrier frequencies are unknown, as well as low rate sampling strategies for other interesting analog models. The insights we gathered so far hint that analog preprocessing is an advantageous route towards developing efficient sub-Nyquist strategies.

——— *Union of Subspaces* ———

**Motivation**

Demodulation, a classic sub-Nyquist strategy, assumes an input signal which lies in certain intervals within the Nyquist range. But, what if the input signal is not limited to a predefined frequency support, or even worse if it spans the entire Nyquist range – can we still reduce the sampling rate below Nyquist? Perhaps surprising, we shall see in the sequel that the answer is affirmative, provided that the input has additional structure we can exploit. Figure 8 illustrates two such scenarios.

Consider for example the scenario of a multiband input $x(t)$ with unknown spectral support, consisting of $N$ frequency bands of individual widths no greater than $B$ Hz. In contrast to the classic setup, the carrier frequencies $f_i$ are unknown, and we are interested in sampling such multiband inputs with transmissions located anywhere below $f_{\max}$. At first sight, it may seem that sampling at the Nyquist rate $f_{\text{NYQ}} = 2f_{\max}$ is necessary, since every frequency interval below $f_{\max}$ appears in the support of some multiband $x(t)$. On the other hand, since each specific $x(t)$ in this model has structure – it fills only a portion of the Nyquist range (only $NB$ Hz) – we intuitively expect to be able to reduce the sampling rate below $f_{\text{NYQ}}$.

Another interesting problem is sampling of signals which consist of several echoes of a known pulse shape, where the delays and attenuations are a-priori unknown. Mathematically,

$$x(t) = \sum_{\ell=1}^{L} a_\ell \, h(t - t_\ell), \quad t \in [0, \tau], \tag{11}$$

for some given pulse shape $h(t)$ and unknown $t_\ell, a_\ell$. Signals of this type belong to the broader family of FRI signals, originally introduced by Vetterli et al. in [27], [28]. Echoes are encountered, for example, in multipath fading communication channels. The transmitter can assist the receiver in channel identification by sending a short probing pulse $h(t)$, based on which the receiver can resolve the fading delays $t_\ell$ and use this information to decode subsequent information messages. In radar applications, inputs of the form (11) are prevalent, where the delays $t_\ell$ correspond to the unknown locations of targets in space, while the amplitudes $a_\ell$ encode Doppler shifts indicating target speeds. Medical imaging techniques, *e.g.,* ultrasound, record signals that are structured according to (11) when probing density changes in human tissue. Underwater acoustics also conform with (11). The common denominator of these applications is that $h(t)$ is a short pulse in time, so that the bandwidth of $h(t)$, and consequently that of $x(t)$, spans a large Nyquist range. Nonetheless, given the structure (11), we can intuitively expect to determine $x(t)$ from samples at the rate of innovation, namely $2L$ samples per $\tau$, which counts the actual number of unknowns, $t_\ell, a_\ell$, $1 \leq \ell \leq L$ in every interval.

These examples hint at a more general notion of sub-Nyquist sampling, in which the underlying signal structure is utilized to reduce acquisition rate below the apparent input bandwidth. As a special case, this notion includes the classic settings of structure given by a predefined frequency support. To capture more general structures, we present next the union of subspace (UoS) model, originally proposed by Lu and Do in [29].

**Mathematical framework**

Denote by $x(t)$ an analog signal in the Hilbert space $\mathcal{H} = L_2(\mathbb{R})$, which lies in a parameterized family of subspaces

$$x(t) \in \mathcal{U} \triangleq \bigcup_{\lambda \in \Lambda} \mathcal{A}_\lambda, \tag{12}$$

where $\Lambda$ is an index set, and each individual $\mathcal{A}_\lambda$ is a subspace of $\mathcal{H}$. The key property of the UoS model (12) is that the input $x(t)$ resides within $\mathcal{A}_{\lambda^*}$ for some $\lambda^* \in \Lambda$, but a-priori, the exact subspace index $\lambda^*$ is unknown. We define the dimension (or bandwidth) of $\mathcal{U}$ as the dimension of its affine hull $\Sigma$, namely the space of all linear combinations of $x(t) \in \mathcal{U}$. Typically, the union $\mathcal{U}$ has dimension that is relatively high compared with those of the individual subspaces $\mathcal{A}_\lambda$.

Multiband signals with unknown carriers $f_i$ can be described by (12), where each $\mathcal{A}_\lambda$ corre-sponds to signals with specific carrier positions and the union is taken over all possible $f_i \in [0, f_{\max}]$. In this case, each $\mathcal{A}_\lambda$ has effective bandwidth $NB$, whereas the union $\mathcal{U}$ has $f_{\max}$ bandwidth, as follows from the definition of $\Sigma$. Similarly, echoes with unknown time-delays of the form (11) correspond to $L$-dimensional subspaces $\mathcal{A}_\lambda$ that capture the amplitudes $a_\ell$. A union over all possible delays $t_l \in [0, \tau]$ provides an efficient way to group these infinitely-many subspaces to a single set $\mathcal{U}$. The large bandwidth of $h(t)$ results in $\mathcal{U}$ with a high Nyquist bandwidth.

Union modeling sheds new light on sampling below the Nyquist rate. Sub-Nyquist in the union setting, conceptually, consists of two layers of rate reduction: from the dimensions of $\mathcal{U}$ to that of the individual subspaces $\mathcal{A}_\lambda$, and then, further reduction within the scope of a single subspace until reaching its effective bandwidth (rather than twice its highest frequency component). The second layer is essentially what is treated in the classic works surveyed earlier, which considered a single subspace defined according to a given spectral support. Eventually, the tricky part is how to design sampling strategies that combine these reduction steps and achieve the minimal rate by one conversion stage. Box 2 expands on the challenges of sampling union sets.

The model (12) can be categorized to four types, according to the cardinality of $\Lambda$ (finite or infinite) and the dimensions of the individual subspaces $\mathcal{A}_\lambda$ (finite or infinite). In the next sections, we review sub-Nyquist sampling methods for several prototype union models (categorized hereafter by the dimensions pair $\lambda - \mathcal{A}_\lambda$, where "F" abbreviates finite):

- multiband with unknown carrier positions (type $F - \infty$),
- variants of FRI models (cover two union types: $\infty - F$ and $\infty - \infty$), and
- a sparse sum of harmonic sinusoids (type $F - F$).

A solution for sampling and reconstruction was developed in [30] for more general $F - F$ union structures. A special case of the $F - F$ case is the sparsity model underlying compressed sensing [31], [32]. In this review, however, our prime focus is analog signals which exhibit infiniteness in either $\Lambda$ or $\mathcal{A}_\Lambda$. A more detailed treatment of the general union setting can be found in [33], [34].

---

**Box 2. Generalized Sampling in Union of Subspaces**

Generalized sampling theory extends upon pointwise acquisition by viewing the measurements as inner products [3]–[6], [35],

$$y[n] = \langle x(t), s_n(t) \rangle, \quad n \in \mathbb{Z}, \tag{13}$$

between an input signal $x(t)$ and a set of sampling functions $s_n(t)$. Geometrically, the sample sequence $y[n]$ is obtained by projecting $x(t)$ onto the space

$$\mathcal{S} = \mathrm{span}\{s_n(t) \,|\, n \in \mathbb{Z}\}. \tag{14}$$

A special case is of a shift-invariant space $\mathcal{S}$ spanned by $s_n(t) = s(t-nT)$ for some generator function $s(t)$ [5]. In this scenario, (13) amounts to filtering $x(t)$ by $s(-t)$ and taking pointwise samples of the output every $T$ seconds. Traditional pointwise acquisition $y[n] = x(nT)$ corresponds to a shift-invariant $\mathcal{S}$ with a Dirac generator $s(t) = \delta(t)$. Multichannel sampling schemes correspond to a shift-invariant space $\mathcal{S}$ spanned by shifts of multiple generators [36], [37].

Theory and applications of subspace sampling were widely studied over the years. If $x(t)$ resides within a subspace $\mathcal{A} \subseteq \mathcal{H}$ of an ambient Hilbert space $\mathcal{H}$, then the samples (13) determine the input whenever the orthogonal complement $\mathcal{A}^\perp$ satisfies a direct sum condition [6]

$$\mathcal{A}^\perp \oplus \mathcal{S} = \mathcal{H}. \tag{15}$$

Reconstruction is obtained by an oblique projection [6]. Roughly speaking, in noiseless settings, perfect recovery is possible whenever the angle $\theta$ between the subspaces $\mathcal{A}, \mathcal{S}$ is different than $90°$, and robustness to noise increases as $\theta$ tends to zero.

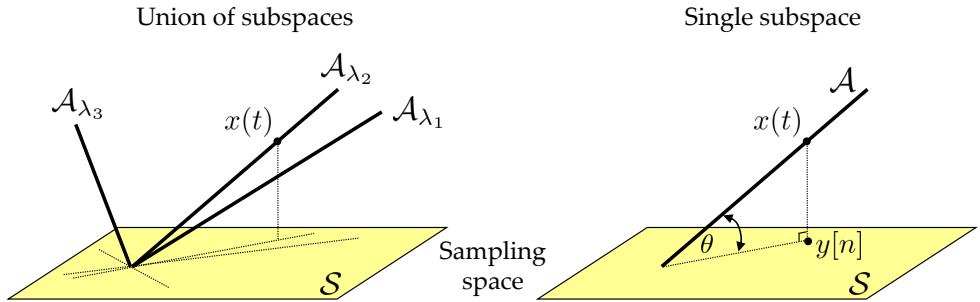

When $x(t)$ lies in a union of subspaces (12), both theory and practice become more intricate. For instance, even if the angles between $\mathcal{S}$ and each of the subspaces $\mathcal{A}_\lambda$ are sufficiently small, the samples may not determine the input if several subspaces are too close to each other; see the illustration. Recent studies [29] have shown that (13) is stably invertible if (and only if) there exist constants $0 < \alpha < \beta < \infty$ such that

$$\alpha \|x_1(t) - x_2(t)\|^2_\mathcal{H} \leq \|y_1[n] - y_2[n]\|^2_{l_2} \leq \beta \|x_1(t) - x_2(t)\|^2_\mathcal{H}, \tag{16}$$

for every signals $x_1(t), x_2(t) \in \mathcal{A}_{\lambda_1} + \mathcal{A}_{\lambda_2}$ and for all possible pairs of $\lambda_1, \lambda_2$. In practice, sampling methods for specific union applications use certain hardware constraints to hint at preferred selections of stable sampling functions $s_n(t)$; see for example [20], [27], [38]–[41] and other UoS methods surveyed in this review.

——— *Multiband Signals with Unknown Carrier Frequencies* ———

**Union modeling**

A description of a multiband union can be obtained by letting $\lambda = \{f_i\}$, so that each choice of carrier positions $f_i$ determines a single subspace $\mathcal{A}_\lambda$ in $\mathcal{U}$. In principle, $f_i$ lies in the continuum $f_i \in [0, f_{\max}]$, resulting in union type $\infty - \infty$ containing infinitely many subspaces. In the setup we describe below a different viewpoint is used to treat the multiband model as a finite union of bandpass subspaces (type $F - \infty$), termed spectrum slices [20], [42].

In this viewpoint, the Nyquist range $[-f_{\max}, f_{\max}]$ is conceptually divided into $M = 2L + 1$ consecutive, non-overlapping, slices of individual widths $f_p$, such that $M f_p \geq 2 f_{\max}$. Each spectrum slice represents a single bandpass subspace. By choosing $f_p \geq B$, we ensure that no more than $2N$ spectrum slices are active, namely contain signal energy. In this setting, there are $\binom{M}{2N}$ subspaces in $\mathcal{U}$. Dividing the spectrum to slices is only a conceptual step, which assumes no specific displacement with respect to the band positions. The advantage of this viewpoint is that switching to union type $F - \infty$ simplifies the digital reconstruction algorithms, while preserving a degree of infiniteness in the dimensions of each individual subspace $\mathcal{A}_\lambda$.

**Semi- and fully-blind pointwise approaches**

Earlier approaches for treating multiband signals with unknown carriers were semi-blind: a sampler design independent of band positions combined with a reconstruction algorithm that requires exact support knowledge. Herley et al. [43] and Bresler et al. [44], [45] studied multi-coset sampling, a PNS grid that is a subset of the Nyquist grid, and proved that the grid points can be selected independently of the band positions. The reconstruction algorithms in [43], [45] coincide with the non-blind PNS reconstruction algorithm of [23], for time-delays $\phi_l$ chosen on the Nyquist grid. These works approach the Landau rate, namely $NB$ samples/sec. Other techniques targeted the rate $NB$ by imposing alternative constraints on the input spectrum [44].

Recently, the math and algorithms for fully-blind systems were developed in [39], [42], [46]. In this setting, both sampling and reconstruction operate without knowledge of the band positions. A fundamental distinction between non- or semi-blind approaches to fully-blind systems is that the minimal sampling rate increases to $2NB$, as a consequence of recovery which lacks knowledge on the exact spectral support. A more thorough discussion in [42] studies the differences between earlier approaches that were based on subspace modeling and the fully-blind sampling methods [39], [42], [46] that are based on union modeling. Box 3 reviews the theorems underlying this distinction. The fully-blind framework developed in [42], [46] provides reconstruction algorithms that can be combined with various sub-Nyquist sampling strategies: multi-coset in [42], filter-bank followed by uniform sampling in [39] and the modulated wideband converter (MWC) in [20]. In viewing our goal of bridging theory and practice, the Achilles heel of the combination with multi-coset is pointwise acquisition, which enters the Nyquist-rate thru the backdoor of T/H bandwidth. As discussed earlier and outlined in Box 1, pointwise acquisition requires an ADC device with Nyquist-bandwidth T/H circuitry. The filter-bank approach is part of a general framework developed in [39] for treating analog signals lying in a sparse-sum of shift-invariant (SI) subspaces, which includes multiband with unknown carriers as a special case. The filters and ADCs, however, also require Nyquist-rate bandwidth, in this setting.

In the next section, we describe the MWC strategy, which utilizes the principles of the fully-blind sampling framework, and also results in a hardware-efficient sub-Nyquist strategy that does not suffer from analog bandwidth limitations of T/H technology. In essence, the MWC extends conventional I/Q demodulation to multiband inputs with unknown carriers, and as such it also provides a scalable solution which decouples undesired RF-ADC dependencies. The combination of hardware-efficient sampler with fully-blind reconstruction effectively satisfies the wishlist of Table 1.

---

### Box 3. Universal Bounds on Sub-Nyquist Sampling Rates

Sampling strategies are often compared on the basis of the required sampling rate. It is therefore instructive to compare existing strategies with the lowest sampling rate possible. For instance, the Shannon-Nyquist theorem states (and achieves) the minimal rate $2f_{\max}$ for bandlimited signals. The following results derive the lowest sub-Nyquist sampling rates for spectrally-sparse signals, under either subspace or union of subspace priors.

Consider the case of a subspace model for signals that are supported on a fixed set $\mathcal{I}$ of frequencies:
$$\mathcal{B}_{\mathcal{I}} = \{x(t) \in L^2(\mathbb{R}) \,|\, \operatorname{supp} X(f) \subseteq \mathcal{I}\}. \tag{17}$$

A grid $R = \{t_n\}$ of time points is called *a sampling set* for $\mathcal{B}_{\mathcal{I}}$ if the sequence of samples

$x_R[n] = x(t_n)$ is stable, namely there exist constants $\alpha > 0$ and $\beta < \infty$ such that:

$$\alpha \|x(t) - y(t)\|_{L_2}^2 \leq \|x_R[n] - y_R[n]\|_{l_2}^2 \leq \beta \|x(t) - y(t)\|_{L_2}^2, \quad \forall x(t), y(t) \in \mathcal{B}_\mathcal{I}. \tag{18}$$

Landau [19] proved that if $R$ is a sampling set for $\mathcal{B}_\mathcal{I}$ then it must have a density

$$D^-(R) \triangleq \lim_{r \to \infty} \inf_{y \in \mathbb{R}} \frac{|R \cap [y, y+r]|}{r} \geq \text{meas}(\mathcal{I}), \tag{19}$$

where $D^-(R)$ is the lower Beurling density and $\text{meas}(\mathcal{I})$ is the Lebesgue measure of $\mathcal{I}$. The numerator in (19) counts the number of points from $R$ in every interval of width $r$ of the real axis. The Beurling density (19) reduces to the usual concept of *average* sampling rate for uniform and periodic nonuniform sampling. Consequently, for multiband signals with $N$ bands of widths $B$, the minimal sampling rate is the sum of the bandwidths $NB$, given a fixed subspace description of known band locations.

A union of subspaces model can describe a more general scenario, in which, a-priori, only the fraction $0 < \Omega < 1$ of the Nyquist bandwidth actually occupied is assumed known but not the band locations:

$$\mathcal{N}_\Omega = \{x(t) \in L^2(\mathbb{R}) \mid \text{meas}(\text{supp}\, X(f)) \leq \Omega f_{\text{NYQ}}\}. \tag{20}$$

A blind sampling set $R$ for $\mathcal{N}_\Omega$ is stable if there exists $\alpha > 0$ and $\beta < \infty$ such that (18) holds with respect to all signals from $\mathcal{N}_\Omega$. A theorem of [42] derived the minimal rate requirement for the set $\mathcal{N}_\Omega$:

$$D^-(R) \geq \min\{2\Omega f_{\text{NYQ}}, f_{\text{NYQ}}\}. \tag{21}$$

This requirement doubles the minimal sampling rate to $2NB$ for multiband signals whose band locations are unknown. It also implies that if the occupation $\Omega > 50\%$, then no rate reduction is possible.

Both minimal rate theorems are universal for pointwise sampling strategies in the sense that for any choice of a grid $R = \{t_n\}$, if the average rate is too low, namely below (19) or (21), then there exist signals whose samples on $R$ are indistinguishable. Note that both results are nonconstructive; they do not hint at a sampling strategy that achieves the minimal rate.

**Modulated wideband converter**

The MWC [20] combines the advantages of RF demodulation and the blind recovery ideas developed in [42], and allows sampling and reconstruction without requiring knowledge of the band locations. To circumvent analog bandwidth issues in the ADCs, an RF front-end mixes the input with periodic waveforms. This operation imitates the effect of delayed undersampling, namely folding the spectrum to baseband with different weights for each frequency interval. In contrast to undersampling (or PNS), aliasing is realized by RF components rather than by taking advantage of the T/H circuitry. In this way, bandwidth requirements are shifted from ADC devices to RF mixing circuitries. The key idea is that periodic mixing serves another goal – both the sampling and reconstruction stages do not require knowledge of the carrier positions.

Before describing the MWC system that is depicted in Fig. 9, we point out several properties of this approach. The system is modular; Sampling is carried out in independent channels, so that the rate can be adjusted to match the requirements of either a traditional subspace model or the more challenging union of subspace prior. It can also scale up to the Nyquist rate to support the standard Shannon-Nyquist bandlimited prior. The reconstruction algorithm that appears in Fig. 12 has several functional blocks: detecting the spectral support thru a computationally light optimization problem, signal recovery and information extraction. Support detection, the heart of this digital algorithm, is carried out whenever the carrier locations vary. The rest of the digital computations are simple and performed in real-time. In addition, the recovery stage outputs baseband samples of $I(t), Q(t)$. This enables a seamless interface to existing DSP algorithms with sub-Nyquist processing rates, as could have been obtained by classic demodulation had the carriers $f_i$ been known. We now elaborate on each part of this strategy.

**Sub-Nyquist sampling scheme**

The conversion from analog to digital consists of a front-end of $m$ channels, as depicted in Fig. 9. In the $i$th channel, $x(t)$ is multiplied by a periodic waveform $p_i(t)$ with period $T_p = 1/f_p$, lowpass filtered by $h(t)$, and then sampled at rate $f_s = 1/T_s$. The figure lists basic and advanced configurations. To simplify, we concentrate on the theory underlying the basic version, in which the sampling interval $T_s$ equals the aliasing period $T_p$, each channel samples at rate $f_s \geq B$ and the number of hardware branches $m \geq 2N$, so that the total sampling rate can be as low as $2NB$. These choices stem from necessary and sufficient conditions derived in [20] on the required sampling rate $mf_s$ to allow perfect reconstruction. If the input's spectral support is known, then the same conditions translate to a similar parameter choice with half the number of channels, resulting in a total sampling rate as low as $NB$. Thus, although the MWC does not take pointwise values of $x(t)$, its optimal sampling rate coincides with the lowest possible rates by pointwise strategies, which are discussed in Box 3. Advanced configurations enable additional hardware savings by collapsing the number of branches $m$ by a factor of $q$ at the expense of increasing the sampling rate of each channel by the same factor, ultimately enabling a single-channel sampling system [20].

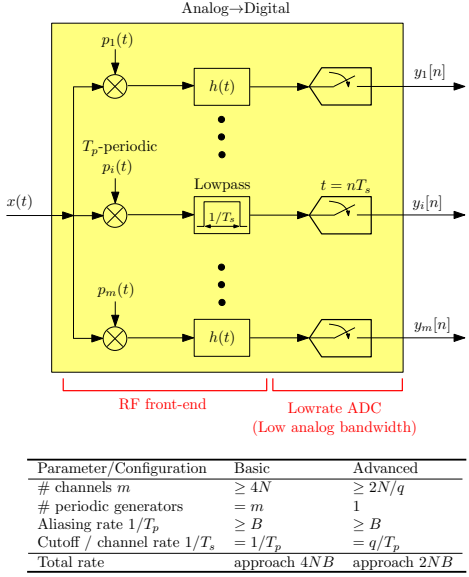
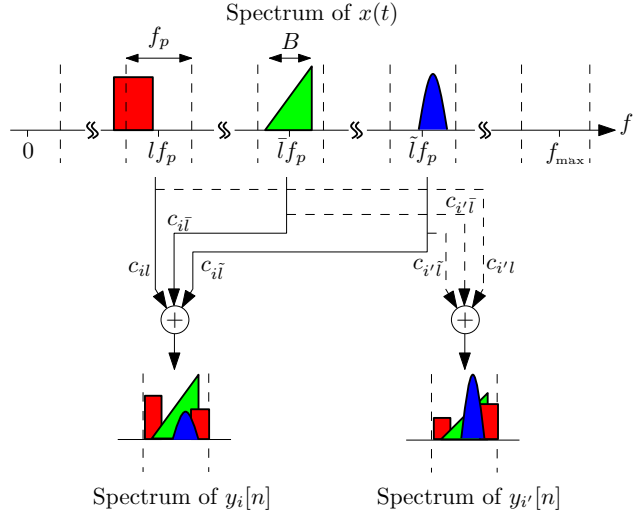

**Fig. 9:** Block diagram of the modulated wideband converter [20]. The table at the bottom lists the parameters choice of the basic and advanced MWC configurations. Adapted from [47], ©[2011] IET.

**Fig. 10:** The spectrum slices from $x(t)$ are overlayed in the spectrum of the output sequences $y_i[n]$. In the example, channels $i$ and $i'$ realize different linear combinations of the spectrum slices centered around $lf_p, \bar{l}f_p, \tilde{l}f_p$. For simplicity, the aliasing of the negative frequencies is not drawn. Adapted from [47], ©[2011] IET.

This technique is also briefly reviewed in the next subsection.

The choice of periodic waveforms $p_i(t)$ becomes clear once analyzing the effect of periodic mixing. Each $p_i(t)$ is periodic, and thus has a Fourier expansion

$$p_i(t) = \sum_{l=-\infty}^{\infty} c_{il} e^{j2\pi f_p l t}. \tag{22}$$

Denote by $z_l[n]$ the sequence that would have been obtained by mixing $x(t)$ with $e^{j2\pi f_p l t}$, filtering by $h(t)$ and sampling every $T$ seconds. By superposition, mixing $x(t)$ by the sum in (22) outputs $y_i[n]$ which is a linear combination of the $z_l[n]$ sequences according to the Fourier coefficients $c_{il}$ of $p_i(t)$. Fig. 10 visualizes the effect of mixing with periodic waveforms, where each sequence $z_l[n]$ corresponds to a spectrum slice of $x(t)$ positioned around $lf_p$. Mathematically, the analog mixture boils down to the linear system [20]

$$\mathbf{y}[n] = \mathbf{C}\mathbf{z}[n], \tag{23}$$

where the vector $\mathbf{y}[n] = [y_1[n], \cdots, y_m[n]]^T$ collects the measurements at $t = nT_s$. The matrix $\mathbf{C}$ contains the coefficients $c_{il}$ and $\mathbf{z}[n]$ rearranges the values of $z_l[n]$ in vector form.

To enable aliasing of spectrum slices up to the maximal frequency $f_{\max}$, the periodic functions $p_i(t)$ need to have Fourier coefficients $c_{il}$ with non-negligible amplitudes for all $-L \leq l \leq L$, such that $Lf_p \geq f_{\max}$. In principle, every periodic function with high-speed transitions within the period $T_p$ can be appropriate. One possible choice for $p_i(t)$ is a sign-alternating function, with $M = 2L + 1$ sign intervals within the period $T_p$ [20]. Popular binary patterns, *e.g.,* the Gold or Kasami sequences, are especially suitable for the MWC [38].

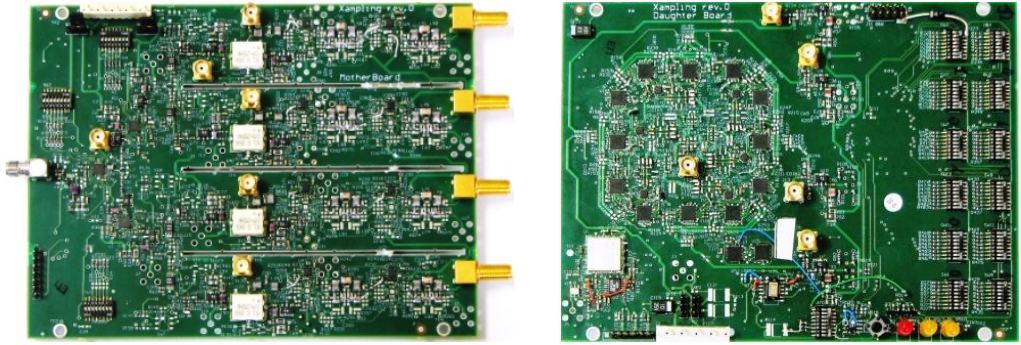

**Fig. 11:** A hardware realization of the MWC consisting of two circuit boards. The left pane implements $m = 4$ sampling channels, whereas the right pane provides four sign-alternating periodic waveforms of length $M = 108$, derived from a single shift-register. Adapted from [47], ©[2011] IET.

**Hardware-efficient realization**

A board-level hardware prototype of the MWC is reported in [47]. The hardware specifications cover 2 GHz Nyquist-rate inputs with spectral occupation up to $NB = 120$ MHz. The sub-Nyquist rate is 280 MHz. Photos of the hardware appear in Fig. 11.

In order to reduce the number of analog components, the hardware realization incorporates an advanced MWC configuration [20]. In this version

- a collapsing factor $q = 3$ results in $m = 4$ hardware branches with individual sampling rates $1/T_s = 70$ MHz; and
- a single shift-register generates periodic waveforms for all hardware branches.

Further technical details on this representative hardware exceed the level of practice we are interested in here, though we emphasize below a few conclusions that connect back to the theory.

The Nyquist burden always manifests itself in some part of the design. For example, in pointwise methods, implementation requires ADC devices with Nyquist-rate front-end bandwidth. In other approaches [41], [48], which we discuss in the sequel, the computational loads scale with the Nyquist rate, so that an input with 1 MHz Nyquist rate may require solving linear systems with 1 million unknowns. Example hardware realizations of these techniques [49] treat signals with Nyquist rate up to 800 kHz. The MWC shifts the Nyquist burden to an analog RF preprocessing stage that precedes the ADC devices. The motivation behind this choice is to enable capturing the largest possible range of input signals, since, in principle, when the same technology is used by the source and sampler, this range is maximal. In particular, as wideband multiband signals are often generated by RF sources, the MWC framework can treat an input range that scales with any advance in RF technology.

While this explains the choice of RF preprocessing, the actual sub-Nyquist circuit design can be quite challenging and call for nonordinary solutions. To give a taste of circuit challenges, we briefly consider two design problems that are studied in detail in [47]. Low cost analog mixers are typically specified for a pure sinusoid in their oscillator port, whereas the periodic mixing requires simultaneous mixing with the many sinusoids comprising $p_i(t)$, which creates nonlinear distortions and complicates the gain selections along the RF path. In [47], special power circuitries that are tailored for periodic mixing were inserted before and after the mixer. Another circuit challenge pertains to generating $p_i(t)$ with 2 GHz alternation rates. The strict timing constraints involved in this logic are eliminated in [47] by operating commercial devices beyond their datasheet specifications.

Going back to a high-level practical viewpoint, besides matching the source and sampler technology and addressing circuit challenges, an important point is to verify that the recovery algorithms do not limit the input range through constraints they impose on the hardware. In the MWC case, periodicity of the waveforms $p_i(t)$ is important since it creates the aliasing effect with the Fourier coefficients $c_{il}$ in (22). The hardware implementation and experiments in [47] demonstrate that the appearance in time of $p_i(t)$ is irrelevant as long as periodicity is maintained[1]. This property is crucial, since precise sign alternations at speeds of 2 GHz are difficult to maintain, whereas simple hardware wirings ensure that $p_i(t) = p_i(t + T_p)$ for every $t \in \mathbb{R}$. The recent work [50] provides digital compensation for nonflat frequency response of $h(t)$, assuming slight oversampling to accommodate possible design imperfections, similarly to oversampling solutions in Shannon-Nyquist sampling.

Noise is inevitable in practical measurement devices. A common property of many existing sub-Nyquist methods, including PNS sampling, MWC and the methods of [41], [48] is that they aggregate wideband noise from the entire Nyquist range, as a consequence of treating all possible spectral supports. The digital reconstruction algorithm we outline in the next subsection partially compensates for this noise enhancement for PNS/MWC by digital denoising. Another route to noise reduction can be careful design of the sequences $p_i(t)$. However, noise aggregation remains

---

[1]A video recording of hardware experiments and additional documentation for the MWC hardware are available at http://webee.technion.ac.il/Sites/People/YoninaEldar/Info/hardware.html. Relevant software packages are available at http://webee.technion.ac.il/Sites/People/YoninaEldar/Info/software.html.

a practical limitation of all current sub-Nyquist techniques.

**Reconstruction algorithm**

The digital reconstruction algorithm encompasses three stages which appear in Fig. 12:

1) A block named continuous-to-finite (CTF) constructs a finite-dimensional frame (or basis) from the samples, from which a small-size optimization problem is formulated. The solution of that program indicates those spectrum slices that contain signal energy. The CTF outputs an index set $S$ of active slices. This block is executed on initialization or when the carrier frequencies change;

2) A single matrix-vector multiplication, per instance of $\mathbf{y}[n]$, recovers the sequences $z_l[n]$ containing signal energy, as indicated by $l \in S$; and

3) A digital algorithm estimates $f_i$ and (samples of) the baseband signals $I(t), Q(t)$ of each information band.

In addition to DSP, analog recovery of $x(t)$ is obtained by remodulating the quadrature signals $I(t), Q(t)$ on the estimated carriers $f_i$ according to (3). Analog back-end employs customary components, DACs and modulators, to recover $x(t)$.

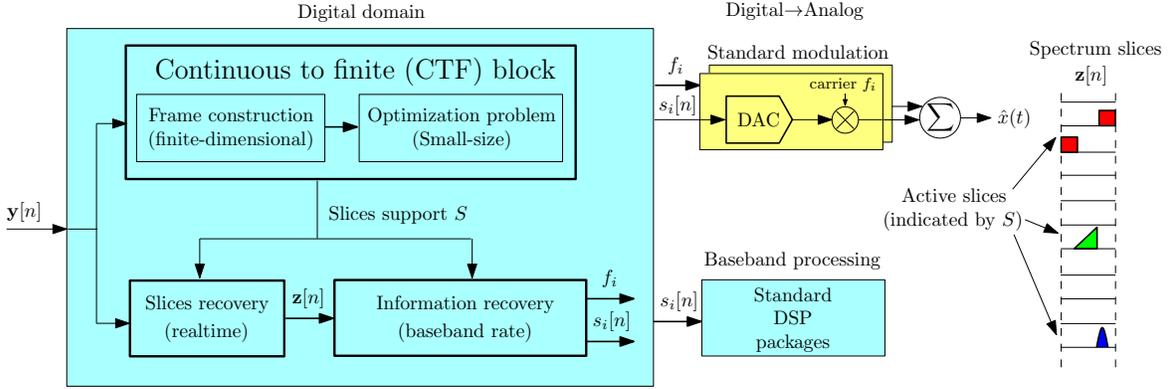

**Fig. 12:** Block diagram of recovery and processing stages of the modulated wideband converter.

To understand the recovery flow, we begin with the linear system (23). Due to the sub-Nyquist setup, the matrix $\mathbf{C}$ in (23) has dimension $m \times M$, such that $m < M$. In other words, $\mathbf{C}$ is rectangular and (23) has less equations than the dimension $M$ of the unknown $\mathbf{z}[n]$. Fortunately, the multiband prior in accordance with the choice $f_p \geq B$ ensures that at most $2N$ sequences $z_l[n]$ contains signal energy [20]. Therefore, for every time point $n$, the unknown $\mathbf{z}[n]$ is sparse with no more than $2N$ nonzero values. Solving for a sparse vector solution of an underdetermined system of equations has been widely studied in the literature of compressed sensing (CS). Box 4 summarizes relevant CS theorems and algorithms.

Recovery of $\mathbf{z}[n]$ using any of the existing sparse recovery techniques is inefficient, since the sparsest solution $\mathbf{z}[n]$, even if obtained by a polynomial-time CS technique, is computed independently for every $n$. Instead, the CTF method that was developed in [42], [46] exploits the fact that the bands occupy continuous spectral intervals. This analog continuity boils down to $\mathbf{z}[n]$ having a common nonzero location set $S$ over time. To take advantage of this joint sparsity, the CTF builds a frame (or a basis) from the measurements using, for example,

$$\mathbf{y}[n] \xrightarrow{\text{Frame construct}} \mathbf{Q} = \sum_n \mathbf{y}[n]\mathbf{y}^H[n] \xrightarrow{\text{Decompose}} \mathbf{Q} = \mathbf{V}\mathbf{V}^H, \qquad (24)$$

where the (optional) decomposition allows to combat noise. The finite dimensional system

$$\mathbf{V} = \mathbf{C}\mathbf{U}, \qquad (25)$$

is then solved for the sparsest matrix $\mathbf{U}$ with minimal number of nonidentically zero rows; example techniques are referenced in Box 4. The important observation is that the indices of the nonzero rows in $\mathbf{U}$, denoted by the set $S$, coincide with the locations of the spectrum slices that contain signal energy [42]. This property holds for any choice of matrix $\mathbf{V}$ in (25) whose columns span the measurements space $\{\mathbf{y}[n]\}$. The CTF effectively locates the signal energy at a spectral resolution of $f_p$. Once $S$ is found, $\mathbf{z}[n]$ are recovered by a matrix-vector multiplication; see (30) in Box 4. Since all CTF operations are executed only once (or when carrier frequencies change), in steady-state, the reconstruction runs in real-time, namely a single matrix-vector multiplication (30) per measurement $\mathbf{y}[n]$.

**Sub-Nyquist baseband processing**

Software packages for DSP expect baseband inputs, namely the information signals $I(t), Q(t)$ of (3), or equivalently their uniform samples at the narrowband rate. These inputs are obtained by classic demodulation when the carrier frequencies are known. A digital algorithm developed in [51] translates the sequences $\mathbf{z}[n]$ to the desired DSP format with only lowrate computations, enabling smooth interfacing with existing DSP software packages.

The input to the algorithm are the sequences $\mathbf{z}[n]$ corresponding to the spectrum slices of $x(t)$. In general, as depicted in Fig. 13, a spectrum slice may contain more than a single information band. The energy of a band of interest may also split between adjacent slices. To correct for these two effects, the algorithm performs the following actions:

1) Refine the coarse support estimate $S$ to the actual band edges, using power spectral density estimation;
2) Separate bands occupying the same slice to distinct sequences $r_i[n]$. Stitch together energy

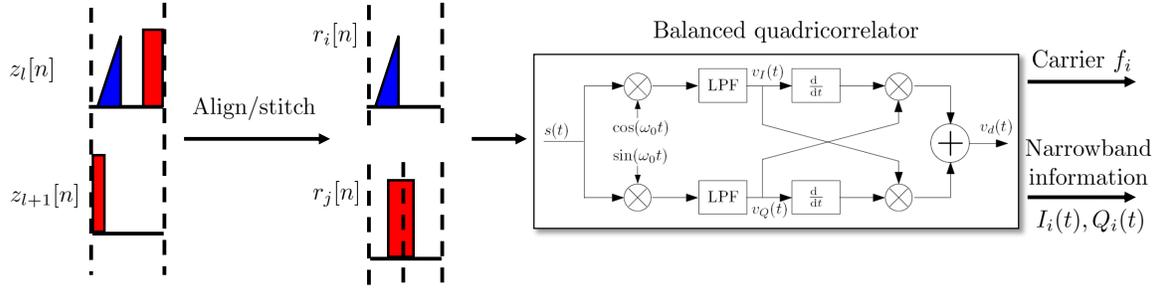

**Fig. 13:** The flow of information extraction begins with detecting the band edges. The slices are filtered, aligned and stitched appropriately to construct distinct quadrature sequences $r_i[n]$ per information band. The balanced quadricorrelator finds the carrier $f_i$ and extracts the narrowband information signals.

that was split between adjacent slices; and

3) Apply a common carrier recovery technique, the balanced quadricorrelator [52], on $r_i[n]$. This step estimates the carrier frequencies $f_i$ and outputs uniform samples of the narrowband signals $I(t), Q(t)$.

The baseband processing algorithm renders the MWC compliant with the high-level architecture of Fig. 2 depicted in the introduction. The digital computations of the MWC (CTF, spectrum slices recovery and baseband processing) lie within the digital core that enables DSP and assist continuous reconstruction.

---

**Box 4. Sparse Solutions of Underdetermined Linear Systems**

A famous riddle reads as follows: "you are given a balanced scale and 12 coins, 1 of which is counterfeit. The counterfeit weighs less or more than the other coins. Determine the counterfeit in 3 weighings, and whether it is heavier or lighter". This riddle captures the essence of sparsity priors. Whilst there are multiple unknowns (the weights of the 12 coins), far fewer measurements (only 3) are required to determine low-dimensional information of interest (the relative weight of the counterfeit coin). Several "12 coins" solutions (widely available online) are based on three rounds of comparing weights of two groups of four coins each, followed by a sort of combinatorial logic that indicates the counterfeit coin.

Sparse solutions of underdetermined linear systems extend the principle underlying the above riddle. The influential works by Donoho [31] and Candès et al. [32] paved the way to compressed sensing (CS), an emerging field in which problems of this type are widely studied. Mathematically, consider the linear system

$$\mathbf{y} = \mathbf{C}\mathbf{z}, \tag{26}$$

with the $m \times M$ matrix $\mathbf{C}$ having fewer rows than columns, *i.e.,* $m < M$. Since $\mathbf{C}$ has a nontrivial null space, there are infinitely many candidates $\mathbf{z}$ satisfying (26). The goal of CS is to find a sparse $\mathbf{z}$ among these solutions, namely a vector $\mathbf{z}$ that contains only few nonzero entries. A basic result in the field [53] shows that (26) has a unique sparse solution if

$$\|\mathbf{z}\|_0 < \frac{1}{2}\left(1 + \frac{1}{\mu}\right), \quad \mu \stackrel{\triangle}{=} \max_{i \neq j} \frac{\langle \mathbf{C}_i, \mathbf{C}_j \rangle}{\|\mathbf{C}_i\| \|\mathbf{C}_j\|}, \qquad (27)$$

where $\|\mathbf{z}\|_0$ counts the number of nonzeros in $\mathbf{z}$, and $\|\mathbf{C}_i\|$ denotes the Euclidian norm of the $i$th column $\mathbf{C}_i$. The unique sparse solution can be found via the minimization program,

$$\min_{\mathbf{z}} \|\mathbf{z}\|_0 \quad \text{s.t.} \quad \mathbf{y} = \mathbf{C}\mathbf{z}. \qquad (28)$$

Similar to the riddle, program (28) is NP-hard with combinatorial complexity.

The CS literature provides polynomial-time techniques for sparse recovery, which coincide with the sparse $\mathbf{z}$ under various conditions on the matrix $\mathbf{C}$. A popular alternative to (28) is solving the convex program

$$\min_{\mathbf{z}} \|\mathbf{z}\|_1 \quad \text{s.t.} \quad \mathbf{y} = \mathbf{C}\mathbf{z}, \qquad (29)$$

where the norm $\|\mathbf{z}\|_1$ sums the magnitudes of the entries. Convex variants of (29) include penalizing terms that account for additive noise. Another approach to sparse recovery are greedy algorithms, which iteratively recover the nonzero locations. For example, orthogonal matching pursuit (OMP) [54] iteratively identifies a single support index. A residual vector $\mathbf{r}$ contains the part of $\mathbf{y}$ that is not spanned by the currently recovered index set. In OMP, an orthogonal projection $\mathbf{P}_S \mathbf{y}$ is computed in every iteration, as described in the figure below. Various greedy approaches are modifications of the main OMP steps.

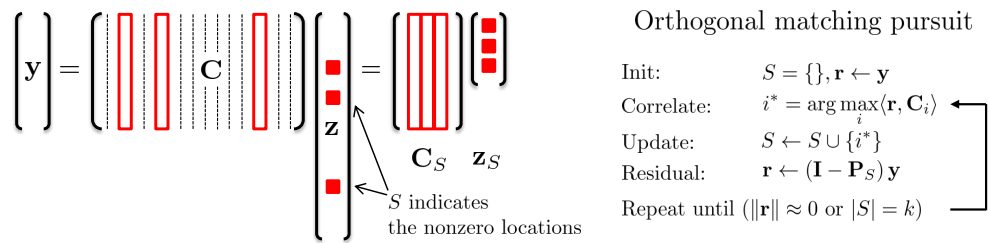

The procedure repeats until the location set $S$ reaches a predefined cardinality or when the residual $\mathbf{r}$ is sufficiently small. Upon termination, the nonzero values $\mathbf{z}_S$ are computed by

> pseudo-inversion of the relevant column subset $\mathbf{C}_S$
>
> $$\mathbf{z}_S = \mathbf{C}_S^\dagger \mathbf{y} = (\mathbf{C}_S^T \mathbf{C}_S)^{-1} \mathbf{C}_S^T \mathbf{y}. \qquad (30)$$
>
> Convex and greedy methods have also been proposed to account for joint sparsity, in which case the unknown is a matrix $\mathbf{Z}$ having only a few nonidentically zero rows [30], [46], [55]–[59]. A special issue of the *Signal Processing Magazine* from March 2008 and [60] provide a comprehensive review of this field [61].

**Adaptive solutions**

We conclude this section with a brief discussion on a potential adaptive strategy for multiband sampling. An adaptive system may scan the spectrum for the frequencies $f_i$ prior to sampling, and then employ classic solutions, *e.g.,* demodulation or PNS, for the actual conversion to digital. This approach requires a wideband analog spectrum-scanner which can be hardware excessive and time consuming; cf. [51]. During that time, signal acquisition is idle, thereby precluding reconstruction of potentially valuable data. The fact that $f_i$ are unknown a-priori and are learnt while the system is running has additional implications on the hardware. For example adaptive demodulation requires a local oscillator tunable over the entire wideband range, so that it can generate a sinusoid at any identified $f_i$ in $[0, f_{\max}]$. In PNS techniques, the sampling grid needs to be designed at run-time, namely after $f_i$ are determined, as evident from conditions (4)-(8) and Figs. 5 and 6. Nonetheless, where applicable, adaptive solutions may be another venue for sub-Nyquist sampling. A prominent advantage of adaptive demodulation is that only in-band noise enters the system.

<div align="center">——— *Signals with Finite Rate of Innovation* ———</div>

**Periodic time-delay model**

Vetterli et al. [27], [62] coined the FRI terminology for signals that are determined by a finite number $L$ of unknowns, referred to as innovations, per time interval $\tau$. Bandlimited signals, for example, have $L = 1$ innovations per Nyquist interval $\tau = 1/f_{\text{NYQ}}$. The most studied FRI model is that of (11), in which there are $2L$ innovations: unknown delays $t_\ell$ and attenuations $a_\ell$ of $L$ copies of a given pulse shape $h(t)$ [27], [28], [40], [62]–[64]. As outlined earlier, the sub-Nyquist goal in this setting is to determine $x(t)$ from about $2L$ samples per $\tau$, rather than sampling at the high rate that stems from the bandwidth of $h(t)$. In what follows, we consider a simple version of

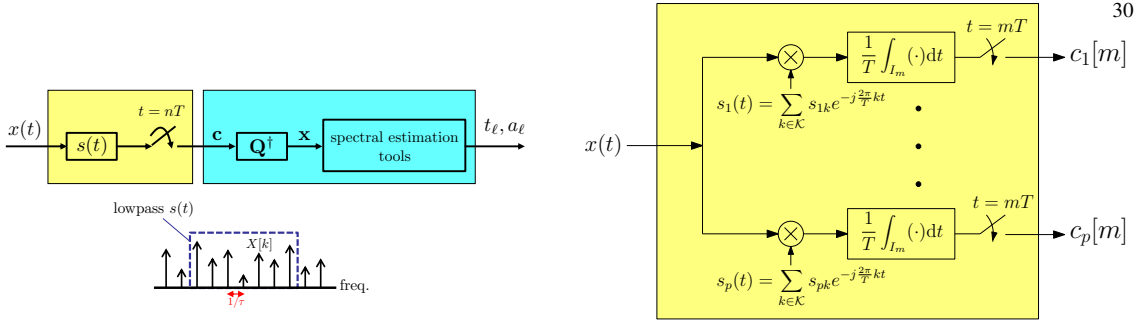

**Fig. 14:** Single and multi-channel sampling schemes for time-delay FRI models.

(11) with a periodic input, $x(t) = x(t+\tau)$, so that the echoes pattern, *i.e.*, $t_\ell$ and $a_\ell$, repeats every $\tau$ seconds. Each possible choice of delays $\{t_\ell\}$ leads to a different $L$-dimensional subspace of signals $\mathcal{A}_\lambda$, spanned by the functions $\{h(t-t_\ell)\}$. Since the delays lie on the continuum $t_\ell \in [0, \tau]$, the model (11) corresponds to an infinite union of finite dimensional subspaces (type $\infty - F$). We first describe the sub-Nyquist principles for this periodic version, and then outline other variants of FRI signals and sampling strategies.

**Sub-Nyquist sampling scheme**

The key enabling sub-Nyquist sampling in the FRI setting is in identifying the connection to a standard problem in signal processing: retrieval of the frequencies and amplitudes of a sum of sinusoids. The Fourier series coefficients $X[k]$ of the periodic pulse stream $x(t)$ can be shown to equal a sum of complex exponentials, with amplitudes $\{a_\ell\}$, and frequencies directly related to the unknown time-delays [27]:

$$X[k] = \frac{1}{\tau}\int_0^\tau x(t)e^{-j2\pi kt/\tau}\mathrm{d}t = \frac{1}{\tau}H(2\pi k/\tau)\sum_{\ell=1}^L a_\ell e^{-j2\pi kt_\ell/\tau}, \qquad (31)$$

where $H(\omega)$ is the Fourier transform of the pulse $h(t)$. Once the coefficients $X[k]$ are known, the delays and amplitudes can be found using standard tools developed in the context of array processing and spectral estimation [27], [65]. Therefore, the goal is to design a sampling scheme from which $X[k]$ can be determined.

Figure 14 depicts two sampling strategies to obtain $X[k]$. In the single-channel version, the input is filtered by $s(t)$ and then sampled uniformly every $T$ seconds. If $s(t)$ is designed to capture a set $\mathbf{x}$ of $M \geq 2L$ consecutive coefficients $X[k]$ and zero out the rest, then the vector $\mathbf{x}$ of Fourier coefficients can be obtained from the samples [63]

$$\mathbf{x} = \mathbf{S}^{-1}\,\mathrm{DFT}\{\mathbf{c}\}, \qquad (32)$$

where **S** is an $M \times M$ diagonal matrix with $k$th entry $S^*(2\pi k/\tau)$ for all $k$ in the filter's passband, and **c** collects $M$ uniform samples in a time duration $\tau$. One way to capture $M$ coefficients $X[k]$ is by choosing a lowpass $s(t)$ with an appropriate cutoff [27]. A more general condition on the sampling kernel $s(t)$ is that its Fourier transform $S(\omega)$ satisfies [63]:

$$S(\omega) = \begin{cases} 0 & \omega = 2\pi k/\tau, k \notin \mathcal{K} \\ \text{nonzero} & \omega = 2\pi k/\tau, k \in \mathcal{K} \\ \text{arbitrary} & \text{otherwise,} \end{cases} \quad (33)$$

where $\mathcal{K}$ is a set of $M \geq 2L$ consecutive indices such that $H\left(\frac{2\pi k}{\tau}\right) \neq 0$ for all $k \in \mathcal{K}$. Practical (real-valued) kernels $s(t)$ have conjugate symmetric transform $S(\omega)$ and thus necessitate selecting odd $M$, in which case the minimal number of samples is $2L + 1$.

A special class of filters satisfying (33) are Sum of Sincs (SoS) in the frequency domain [63], which lead to compactly supported filters in the time domain; this property becomes crucial in other variants of FRI models we survey below. As the name hints, SoS filters are given in the Fourier domain by

$$G(\omega) = \frac{\tau}{\sqrt{2\pi}} \sum_{k \in \mathcal{K}} b_k \operatorname{sinc}\left(\frac{\omega}{2\pi/\tau} - k\right), \quad (34)$$

where $b_k \neq 0$, $k \in \mathcal{K}$. It is easy to see that this class of filters satisfies (33) by construction. Switching to the time domain

$$g(t) = \operatorname{rect}\left(\frac{t}{\tau}\right) \sum_{k \in \mathcal{K}} b_k e^{j2\pi kt/\tau}, \quad (35)$$

which is clearly a time compact filter with support $\tau$. For the special case in which $\mathcal{K} = \{-p, \ldots, p\}$ and $b_k = 1$,

$$g(t) = \operatorname{rect}\left(\frac{t}{\tau}\right) \sum_{k=-p}^{p} e^{j2\pi kt/\tau} = \operatorname{rect}\left(\frac{t}{\tau}\right) D_p(2\pi t/\tau), \quad (36)$$

where $D_p(t)$ denotes the Dirichlet kernel.

An alternative multi-channel sampling system was proposed in [64]. The system, depicted in the right pane of Fig. 14, is conceptually constructed in two steps. First, $M$ analog branches are used to compute $X[k]$ directly from $x(t)$ according to (31): modulation by $e^{-j2\pi kt/\tau}$ and integration over $\tau$. For practical reasons, generating $M$ complex sinusoids at different frequencies can be hardware excessive. Therefore, the second step replaces mixing with individual sinusoids by $x(t)s_i(t)$, with mixing waveforms $s_i(t)$ consisting of a linear combination of $|\mathcal{K}|$ complex sinusoids. The advantage is that $s_i(t)$ can be efficiently generated by proper (lowpass) filtering of periodic waveforms. The periodic waveforms themselves can be generated from a single clock source [47]. Interestingly, the MWC hardware prototype, whose boards appear in Fig. 11, functions as a generic sub-Nyquist platform; it can also be used for reduced-rate sampling of FRI models [66]. In the digital domain, $X[k]$ are computed from samples of the linear mixtures $x(t)s_i(t)$.

**Reconstruction algorithm**

Given a vector $\mathbf{x}$ of coefficients $X[k]$, solving for $t_\ell, a_\ell$ from (31) is tantamount to recovering $L$ frequencies and amplitudes in a sum of complex exponentials. A variety of methods for that problem have been proposed; see [65] for a comprehensive review. Below we outline the annihilating filter method that is used in [27], as it allows recovery from the critical rate of $2L/\tau$.

The key ingredient of the method is a digital filter $A[k]$, whose $z$-transform

$$A(z) = \sum_{k=0}^{L} A[k]z^{-k} = A[0] \prod_{l=1}^{L} \left(1 - e^{-j2\pi t_\ell/\tau} z^{-1}\right) \qquad (37)$$

has zeros at the $L$ fundamental frequencies $e^{j2\pi t_\ell/\tau}$. Convolving $A[k]$ with the coefficients $X[k]$, has an annihilating effect, namely returns zero, since each of the frequencies in $X[k]$ is canceled out by the relevant zero of $A(z)$. The idea is therefore to construct $A[k]$ and then factorize its roots to recover the fundamental frequencies, which imply $t_\ell$. In turn, the amplitudes $a_\ell$ are found by standard linear regression tools. The annihilating filter $A[k]$ is computed from the set of constraints [27], [65]

$$\begin{bmatrix} X[0] & X[-1] & \cdots & X[-L] \\ X[1] & X[0] & \cdots & X[-(L-1)] \\ \vdots & \vdots & \ddots & \vdots \\ X[L] & X[L-1] & \cdots & X[0] \end{bmatrix} \begin{pmatrix} A[0] \\ A[1] \\ \vdots \\ A[L] \end{pmatrix} = \mathbf{0}. \qquad (38)$$

Without loss of generality $A[0] = 1$ (constant scaling does not affect the roots in (37)), so that (38) determines the annihilating filter, and consequently $\{t_\ell\}_{\ell=1}^{L}$, from as low as $2L$ values of $X[k]$. As explained before, a single-channel real-valued kernel $s(t)$ requires a minimal number of samples equal to $M = 2L + 1$.

**Finite-duration FRI models**

While periodic streams are mathematically convenient, finite pulse streams of the form (11) are ubiquitous in real world applications. For example, in ultrasound imaging, there are finitely-many echoes reflected from the tissue. Radar techniques determine target locations based on echoes of a transmitted pulse, where again finitely-many echoes are used. A finite-duration FRI input $x(t)$ coincides with its periodized version $\sum_{k \in \mathbb{Z}} x(t + k\tau)$ on the observation interval $[0, \tau]$. Thus, ultimately, we would like to utilize the previous sampling techniques and algorithms on that interval. The difficulty is, however, that a simple lowpass kernel $s(t)$ has infinite time support, which lasts effectively beyond the time interval $[0, \tau]$, to the point where $x(t)$ differs from its periodized version. A more localized Gaussian sampling kernel was proposed in [27]; however, this method is numerically unstable since the samples are multiplied by a rapidly diverging or decaying exponent. Compactly supported sampling kernels based on splines were studied in [28] for certain classes of pulse shapes. These kernels enable computing moments of the signal rather than its Fourier coefficients, which are then processed in a similar fashion to obtain $t_\ell, a_\ell$.

An alternative approach is to exploit the compact support of SoS filters [63]. Since (35) is compactly supported in time by construction, the values of $x(t)$ beyond the filter support are of no interest. In particular, $x(t)$ may be zero in that range. Therefore, when using SoS filters, periodic and finite-duration FRI models are essentially treated in the same fashion. This approach exhibits superior noise robustness when compared to the Gaussian and spline methods, and can be used for stable reconstruction even for very high values of $L$, e.g., $L = 100$. Potential applications are ultrasound imaging [63], radar [67] and Gabor analysis in the Doppler plane [68]. Multichannel sampling, according to Fig. 14, can be more efficient for implementation since accurate delay elements are avoided. The parallel scheme enjoys similar robustness to noise and allows approaching the minimal innovation rate. It is also applicable in cases of infinite pulse streams, as we discuss next.

**Infinite pulse stream**

The model of (11) can be further extended to an infinite stream case, in which

$$x(t) = \sum_{l \in \mathbb{Z}} a_\ell h(t - t_\ell), \quad t_\ell, a_\ell \in \mathbb{R}. \tag{39}$$

Both [28] and [63] exploit the compact support of their sampling filters, and show that under certain conditions the infinite stream may be divided into a series of finite-duration FRI problems, which are each solved independently using the previous algorithms. Since proper spacing is required between the finite streams in order to ensure up to $L$ pulses within the support of the sampling kernel, the rate is increased beyond minimal. In [28], the rate scales with $L^2$, whereas in [63] the rate requirement is improved to about $6L$, namely a small constant factor from the rate of innovation. A multi-channel approach for the infinite model was first considered for Dirac streams, where a successive chain of integrators allows obtaining moments of the signal [69]. Exponential filters were used in [70] for the same model of Dirac streams. Unfortunately, both methods are sensitive in the presence of noise and for large values of $L$ [64]. A simple sampling and reconstruction scheme consisting of two R-C circuit channels was presented in [71] for the special case where there is no more than one Dirac per sampling period. The system of Fig. 14 can treat a broader class of infinite pulse streams, while being much more stable [64]. It exhibits superior noise robustness over the integrator chain method [69] and allows for more general compactly-supported pulse shapes.

**Sequences of innovations**

A special case of (39) is when the time delays repeat periodically (but not the amplitudes), resulting in

$$x(t) = \sum_{n \in \mathbb{Z}} \sum_{\ell=1}^{L} a_\ell[n] h(t - t_\ell - nT), \quad (40)$$

where $\lambda = \{t_\ell\}_{\ell=1}^{L}$ is a set of unknown time delays contained in the time interval $[0, T]$, $\{a_\ell[n]\}$ are arbitrary bounded energy sequences and $h(t)$ is a known pulse shape. For a given set of delays $\lambda$, each signal of the form (40) lies in a shift-invariant subspace $\mathcal{A}_\lambda$, spanned by $L$ generators $\{h(t - t_\ell)\}_{\ell=1}^{L}$. Since the delays can take on any values in the continuous interval $[0, T]$, the set of all signals of the form (40) constitutes an infinite union of shift-invariant subspaces $|\Lambda| = \infty$. Additionally, since any signal has parameters $\{a_\ell[n]\}_{n \in \mathbb{Z}}$, each of the $\mathcal{A}_\lambda$ subspaces has infinite cardinality, *i.e.*, union type $\infty - \infty$. This model can represent, for example, a time-division multiple access (TDMA) setup, in which $L$ transmitters access a joint channel on predefined time-slots. Due to unknown propagations in the channel, the receiver intercepts symbol streams $a_\ell[n]$ at unknown delays $t_\ell$.

A sampling and reconstruction scheme for signals of the form (40) is depicted in Fig. 15 [40]. The multi-channel scheme has $p$ parallel sampling channels. In each channel, the input signal $x(t)$ is filtered by a band-limited sampling kernel $s_\ell^*(-t)$ with frequency support contained in an interval of width $2\pi p/T$, followed by a uniform sampler at rate $1/T$, thus providing the sampling sequence $c_\ell[n]$. Note that just as in the MWC system, the multiple branches can be collapsed to a single filter whose output is sampled $p$ times faster. The role of the sampling kernels is to smear the pulse in time, prior to low rate sampling.

To recover the signal from the samples, a properly designed digital filter correction bank, whose frequency-domain response is given by $\mathbf{W}^{-1}(e^{j\omega T})$, is applied to the sampling sequences $c_\ell[n]$. The entries of $\mathbf{W}(e^{j\omega T})$ depend on the choice of the sampling kernels $s_\ell^*(-t)$ and pulse shape $h(t)$ by

$$\mathbf{W}\left(e^{j\omega T}\right)_{\ell,m} = \frac{1}{T} S_\ell^*(\omega + 2\pi m/T) H(\omega + 2\pi m/T). \quad (41)$$

The corrected sample vector $\mathbf{d}[n] = [d_1[n], \ldots, d_p[n]]^T$ is related to the unknown amplitude vector $\mathbf{a}[n] = [a_1[n], \ldots, a_L[n]]^T$ by a Vandermonde matrix which depends on the unknown delays $t_\ell$ [40]. Therefore, subspace detection methods, such as the ESPRIT algorithm [72], can be used to recover the delays $\lambda = \{t_1, \ldots, t_L\}$. Once the delays are determined, additional filtering operations are applied on the samples to recover the information sequences $a_\ell[n]$. In particular, referring to Fig. 15, the matrix $\mathbf{D}$ is a diagonal matrix with diagonal elements equal to $e^{-j\omega t_k}$, and $\mathbf{N}(\lambda)$ is a Vandermonde matrix with elements $e^{-j2\pi m t_k/T}$.

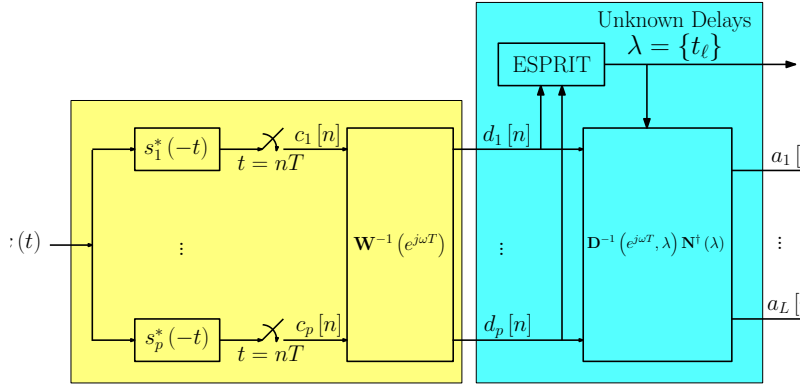

**Fig. 15:** Sampling and reconstruction scheme for signals of the form (40).

In general, the number of sampling channels $p$ required to ensure unique recovery of the delays and sequences using the proposed scheme has to satisfy $p \geq 2L$ [40]. This leads to a minimal sampling rate of $2L/T$. For certain signals, the sampling rate can be reduced even further to $(L+1)/T$ [40]. Evidently, the minimal sampling rate is not related to the Nyquist rate of the pulse $h(t)$. Therefore, for wideband pulse shapes, the reduction in rate can be quite substantial. As an example, consider the setup in [73], used for characterization of ultra-wide band wireless indoor channels. Under this setup, pulses with bandwidth of $W = 1$ GHz are transmitted at a rate of $1/T = 2$MHz. Assuming that there are 10 significant multipath components, this method reduces the sampling rate down to 40MHz compared with the 2GHz Nyquist rate.

**Noise-free vs. noisy FRI models**

The performance of FRI techniques was studied in the literature mainly for noise-free cases. When the continuous-time signal $x(t)$ is contaminated by noise, recovery of the exact signal is no longer possible regardless of the sampling rate. Instead, one may speak of the minimum squared error (MSE) in estimating $x(t)$ from its noisy samples. In this case the rate of innovation $L$ takes on a new meaning as the ratio between the best MSE achievable by any unbiased estimator and the noise variance $\sigma^2$, regardless of the sampling method [74]. This stands in contrast to the noise-free interpretation of $L$ as the minimum sampling rate required for perfect recovery.

In general, the sampling rate which is needed in order to achieve an MSE of $L\sigma^2$ is equal to the rate associated with the affine hull $\Sigma$ of the union set [74]. In some cases, this rate is finite, *e.g.,* in a multiband union, but in many cases the sum covers the entire space $L_2(\mathbb{R})$, *e.g.,* an FRI union with nonbandlimited pulse shape $h(t)$, in which case no finite-rate technique achieves the optimal MSE. This again is quite different from the noise-free case, in which recovery is usually possible at a rate of $2L$, where $L$ is the individual subspace dimension.

A consequence of these results is that oversampling can generally improve estimation performance. Indeed, it can be shown that sampling rates much higher than $L$ are required in certain settings in order to approach the optimal performance. Furthermore, these gains can be substantial: In some cases, oversampling can improve the MSE by several orders of magnitude. These results help explain effects of numerical instability which are sometimes observed in FRI reconstruction. As a rule of thumb, it appears that for union of subspace signals, performance is improved at low rates if most of the unknown parameters identify the position within the subspace $\mathcal{A}_\lambda$, rather than the subspace index $\lambda^*$. Further details on these bounds and recovery performance appear in [74].

——— *Sparse Sum of Harmonic Sinusoids* ———

**Discretized model**

Rapid interest in CS over the last few years has given a major drive to sub-Nyquist sampling. CS focuses on efficiently measuring a discrete signal (vector) $\mathbf{z}$ of length $M$ that has $k < M$ nonzero entries. A measurement vector $\mathbf{y}$ of shorter length, proportional to $k$, is generated by $\mathbf{y} = \mathbf{\Phi z}$, using an underdetermined matrix $\mathbf{\Phi}$. Since $\mathbf{z}$ is sparse, it can be recovered from $\mathbf{y}$, even though $\mathbf{\Phi}$ has less rows than columns. Box 4 elaborates more on the techniques used in CS for sparse vector reconstruction. The CS setup borrows the sub-Nyquist terminology for the finite setting, so as to emphasize that the measurement vector $\mathbf{y}$ is shorter than $\mathbf{z}$.

Although CS is in essence a mathematical theory for measuring finite-length vectors, various researchers applied these ideas to sensing of analog signals by using discretized or finite-dimensional models [41], [75]–[77]. One of the first works in this direction [41] explores CS techniques for sensing a sparse sum of harmonic tones

$$f(t) = \sum_{k=-(W/2-1)}^{W/2} a_k e^{j2\pi kt}, \quad \text{for } t \in [0,1) \tag{42}$$

with at most $k$ nonzero coefficients $a_k$ out of $W$ possible tones. In contrast to FRI models which permit $t_\ell$ to lie on the continuum, the active sinusoids in (42) lie on a uniform grid of frequencies $\{k\Delta\}$ with normalized spacing $\Delta = 1$ (union type F − F).

The random demodulator (RD) senses a sparse harmonic input $f(t)$ by mapping blocks of Nyquist-rate samples to low rate measurements via a binary random combination, as depicted in

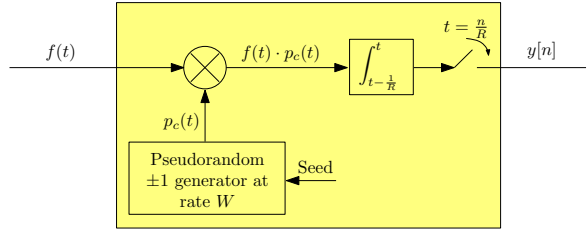

**Fig. 16:** Block diagram of the random demodulator [41].

Fig. 16. The signal $f(t)$ is multiplied by a pseudorandom $\pm 1$ generator alternating at rate $W$, and then integrated and dumped at a constant rate $R < W$. A vector $\mathbf{y}$ collects $R$ consecutive measurements, resulting in the underdetermined system [41]

$$\mathbf{y} = \mathbf{\Phi}\mathbf{f} = \mathbf{\Phi}\,\mathrm{DFT}\{\mathbf{z}\}, \tag{43}$$

where the random sign combinations are the entries of $\mathbf{\Phi}$ and $\mathbf{f}$ corresponds to the values of $f(t)$ on the Nyquist grid (more precisely, the entries of $\mathbf{f}$ are the values that are obtained by integrating-and-dumping $f(t)$ on $1/W$ time intervals). The vector of DFT coefficients $\mathbf{z}$ coincides with $a_k$ due to the time-axis normalization $\Delta = 1$. Using CS recovery algorithms, $\mathbf{z}$ is determined and then $f(t)$ is resynthesized using (42). A bank of RD channels with overlapping integrations was studied in [48].

The RD method is one of the pioneer and innovative attempts to extend CS to analog signals. Underlying this approach is input modeling that relies on finite discretization. Thus, as long as the signal obeys this finite model, as in the case, for example, with harmonic tones (42), extending CS is possible following this strategy. In practice, however, in many applications we are interested in processing and representing an underlying analog signal which is decidedly not finite-dimensional, *e.g.,* multiband or FRI inputs. Applying discretization on analog signals which posses infinite structures can result in large hardware and software complexities, as we discuss in the next subsection.

**Discretization vs. continuous analog modeling**

Transition from analog to digital is one of the tricky parts in any sampling strategy. The approach we have been describing in this review treats analog signals by taking advantage of UoS modeling, where infiniteness enters either through the dimensions of the underlying subspaces $\mathcal{A}_\lambda$, the cardinality of the union $|\Lambda|$, or both (types $\mathrm{F}-\infty$, $\infty-\mathrm{F}$ and $\infty-\infty$, respectively). The sparse harmonic model is, however, exceptional since in this case both $\Lambda$ and $\mathcal{A}_\lambda$ are finite (type $\mathrm{F}-\mathrm{F}$). It is naturally tempting to use finite tools and to avoid the difficulties that come with infinite structures. Theoretically, an analog multiband signal can be approximated to a desired

38|  |  | RD |  | MWC |  |
|---|---|---|---|---|---|
|  | Discretization spacing | $\Delta = 1$ Hz | $\Delta = 100$ Hz |  |  |
| Model | $K$ tones | $300 \cdot 10^6$ | $3 \cdot 10^6$ | $N$ bands | 6 |
|  | out of $Q$ tones | $10 \cdot 10^{10}$ | $10 \cdot 10^8$ | width $B$ | 50 MHz |
| Sampling setup | alternation speed $W$ | 10 GHz | 10 GHz | $m$ channels | 35 |
|  |  |  |  | $M$ Fourier coefficients | 195 |
|  |  |  |  | $f_s$ per channel | 51 MHz |
|  | rate $R$ | 2.9 GHz | 2.9 GHz | total rate | 1.8 GHz |
| Preparation |  |  |  |  |  |
| Collect samples | Num. of samples $N_R$ | $2.9 \cdot 10^9$ | $2.9 \cdot 10^7$ | $2N$ snapshots of $\mathbf{y}[n]$ | $12 \cdot 35 = 420$ |
| Delay | $N_R/R$ | 1 sec | 10msec | $2N/f_s$ | 235nsec |
| CS block |  |  |  |  |  |
| Matrix dimensions | $\mathbf{\Phi F} = N_R \times Q$ | $2.6 \cdot 10^9 \times 10^{10}$ | $2.6 \cdot 10^7 \times 10^8$ | $\mathbf{C} = m \times M$ | $35 \times 195$ |
| Apply matrix | $\mathcal{O}(W \log W)$ |  |  | $\mathcal{O}(mM + M \log M)$ |  |
| Storage | $\mathcal{O}(W)$ |  |  | $\mathcal{O}(mM)$ |  |
| Real-time (fixed support) |  |  |  |  |  |
| Memory length | $N_R$ | $2.9 \cdot 10^9$ | $2.9 \cdot 10^7$ | 1 snapshot of $\mathbf{y}[n]$ | 35 |
| Delay | $N_R/R$ | 1 sec | 10msec | $1/f_s$ | 19.5nsec |
| Mult.-ops. (Millions/sec) | $KN_R/(N_R/R)$ | $8.7 \cdot 10^{11}$ | $8.7 \cdot 10^9$ | $2Nmf_s$ | 21420 |

[Table 2] IMPACT OF DISCRETIZATION ON COMPUTATIONAL LOADS.

precision by a dense grid of discrete tones [41]. However, there is a practical price to pay – the finite dimensions grow arbitrary large; a 1 MHz Nyquist-rate input boils down to a sparse recovery problem with $W = 10^6$ entries in $\mathbf{z}$. In addition, discretization brings forth sensitivity issues and loss of signal resolution as demonstrated in the sequel.

To highlight the issues that result from discretization of general analog models, we consider an example scenario of a wideband signal with $f_{\text{NYQ}} = 10$ GHz, 3 concurrent transmissions and 50 MHz bandwidths around unknown carriers $f_i$. Table 2, quoted from [51], compares various digital complexities of the MWC and an RD system which is applied on a $\Delta$-spaced grid of frequencies for two discretization options $\Delta = 1$ Hz and $\Delta = 100$ Hz. The notation in the table is self-explanatory. It shows that discretization of general continuous inputs results in computational loads that effectively scale with the Nyquist rate of the input, which can sometimes be orders of magnitude higher than the complexity of approaches that directly treat the infinite union structure.

The differences reported in Table 2 stem from attempting to approximate a multiband model by a discrete set of tones, so as to consider inputs with comparable Nyquist bandwidth. At first sight, the signal models and compression techniques used in the MWC and RD seem similar, at least visually. A comprehensive study in [51] examines each system with its own model and compares them in terms of hardware and software complexities and robustness to model mismatches (as also briefly discussed in Box 5). This comparison reveals that in this setting the MWC outperforms the RD, at least in these practical metrics, with a sampler hardware that can be readily implemented with existing analog devices and computationally-light software algorithms. Similar conclusions were reached in [67], where sub-Nyquist radar imaging developed based on union modeling was demonstrated to accomplish accurate target identification and super-resolution capabilities in high signal-to-noise ratio (SNR) environments. In comparison, discretization-based approaches for radar imaging in high SNR were shown to suffer from spectral leakage which degrades identification accuracy and has limited super-resolution capabilities even in noise free settings.

The conclusion we would like to convey is that union modeling provides a convenient mechanism to preserve the inherent infiniteness that many classes of analog signals posses. The infiniteness can enter thru the dimensions of the individual subspaces $\mathcal{A}_\lambda$, the union cardinality $|\Lambda|$, or both. Alternative routes relying on finite models to approximate continuous signals, presumably via discretization, may lead to high computational complexities and strong sensitivities. Nonetheless, there are examples of continuous-time signals that naturally possess finite representations (one such example are trigonometric polynomials). In such situations of an input that is well approximated by a regularized finite model of small size, analog discretization can be beneficial. It is therefore instructive to examine the specific acquisition problem at hand and choose between analog-based sampling to the discretization-based alternative. In either option, applying CS techniques in the digital domain, as part of reconstruction, can bring forward prominent advantages, *i.e.,* provable robustness to noise and widely available off-the-shelf solvers. One potential application of CS is in the context of FRI recovery, where instead of using ESPRIT, MUSIC or annihilating filter for time-delay estimation on the continuum, one can consider discretizing the reconstruction time-axis and using a CS solver to increase the overall noise robustness [78].

In the next section, we summarize our review and discuss the potential of sub-Nyquist strategies to appear in real-world applications.

——— *Summary* ———

**From theory to practice**

We began the review with the Shannon-Nyquist theorem. Undoubtedly, uniform sampling ADC devices are the most common technology in the market. Fig. 17 maps off-the-shelf ADC devices according to their sampling rate. The ADC industry has perpetually followed the Nyquist paradigm – the datasheets of all the devices that are reported in the figure highlight the conversion speed, referring to uniform sampling of the input. The industry is continuously striving to increase the possible uniform conversion rates.

Concluding this review, we would like to focus on multiband inputs and sketch the scenarios that may justify employing a sub-Nyquist solution over the traditional DSP scheme of Fig. 1. Tables 3-4 summarize the sub-Nyquist methods we surveyed earlier. Among the subspace methods

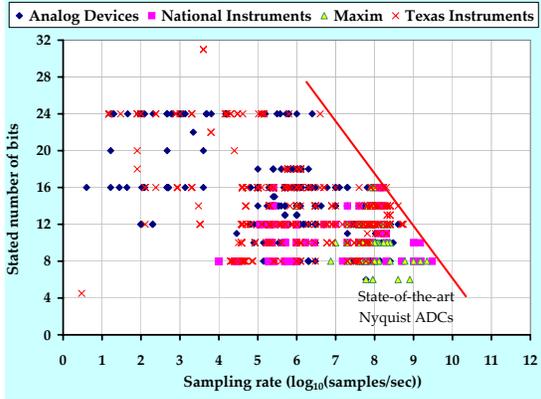

**Fig. 17:** ADC technology – Stated number of bits versus sampling rate. A map of more than 1200 ADC devices from four leading manufacturers, according to online datasheets [47]. Previous mappings from the last decade are reported in [7], [8].

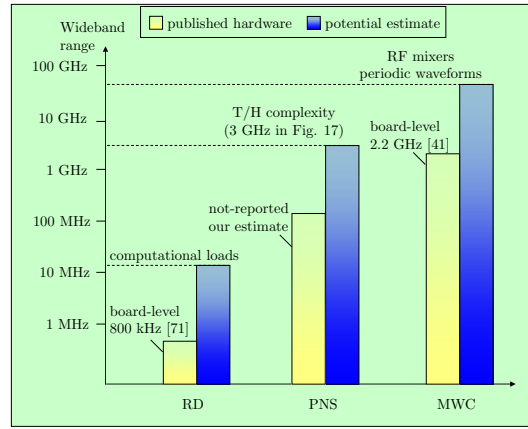

**Fig. 18:** Technology potential of state-of-the-art sub-Nyquist strategies (for multiband inputs).

| | Strategy | Model | Cardinality | | Analog pre-processing | Req. ADC bandwidth | Reconstruction principle | Sub-Nyquist processing | Status | Technology barrier |
|---|---|---|---|---|---|---|---|---|---|---|
| | | | $|\Lambda|$ | $\mathcal{A}_\lambda$ | | | | | | |
| | [Table 3] SUB-NYQUIST STRATEGIES (SPECTRALLY-SPARSE). | | | | | | | | | |
| Classic | Shannon-Nyquist | bandlimited | 1 | $\infty$ | | Nyquist | Interpolation (1) | | Commercial | ADC |
| | Demodulation | multiband | 1 | $\infty$ | I/Q demod. | lowrate | DAC + modulation | ✓ | Commercial | RF |
| | Undersampling [18] | bandpass | 1 | $\infty$ | delay | Nyquist | piecewise filtering | | | ADC (T/H) |
| | PNS [22], [23], [43], [45] | multiband | 1 | $\infty$ | delay | Nyquist | piecewise filtering | | | ADC (T/H) |
| Union of subspaces | PNS [42] | multiband | $M$ | $\infty$ | delay | Nyquist | CTF | | | ADC (T/H) |
| | Filter-bank [39] | sparse-SI | $M$ | $\infty$ | filters | Nyquist | CTF | | | ADC (T/H) |
| | MWC [20] | multiband | $M$ | $\infty$ | periodic-mixing | low | CTF | ✓ | 2.2 GHz board-level [47] | RF |
| | RD [41] | sparse harmonic | $K$ | $W$ | random-sign mixing | low | CS | | 800 kHz board-level [49] | software |

demodulation is already adapted by industry for sampling a multiband input below $f_{\text{NYQ}}$ when the carrier positions are known. Undersampling is also popular to some extent when there is a single band of information, and the maximal frequency $f_u < b$, namely within the available T/H bandwidths of commercial ADC devices. In contrast, although popular in time-interleaved ADCs, PNS was not widely embraced for sub-Nyquist sampling. This situation is perhaps reasoned by

| [Table 4] SUB-NYQUIST STRATEGIES (FINITE RATE OF INNOVATION). | | | |
|---|---|---|---|
| Analog preprocessing | Cardinality | | Reconstruction algorithm |
| | $|\Lambda|$ | $\mathcal{A}_\lambda$ | |
| Lowpass [27], [62] | $\infty$ | $2L$ | annihilating filter |
| Gaussian [27], [62] | $\infty$ | $2L$ | annihilating filter |
| Poly.-/exp.-reproducing kernel [28] | $\infty$ | $2L$ | moments filtering |
| Succ.-integration [69] | $\infty$ | $2L$ | annihilating filter |
| Exp.-filtering [70] | $\infty$ | $2L$ | pole-cancelation filter |
| RC-circuit [71] | $\infty$ | $2$ | closed-form |
| SoS-filtering [63] | $\infty$ | $2L$ | annihilating filter |
| Periodic-mixing [64] | $\infty$ | $2L/\infty$ | annihilating filter |
| Filter-bank [40] | $\infty$ | $\infty$ | MUSIC [79] / ESPRIT [72] |

the fact that the technology barrier of any pointwise method is eventually limited by the analog bandwidth of the T/H stage. Accumulating wideband noise is another drawback of PNS (and the MWC and RD).

When the carrier frequencies are unknown, single subspace methods are not an option anymore. In Fig. 18, we draw the rough potential of three leading sub-Nyquist technologies (for multiband inputs) as we foresee. The MWC approach extends the capabilities of I/Q demodulation by mixing the input with multiple sinusoids at once, probably limited by noise. T/H limitations remain the bottleneck of PNS alternatives [42]. We added the RD to Fig. 18 despite the fact that it treats harmonic inputs rather than narrowband transmissions. The figure reveals that while hardware constraints bound the potential of sampling strategies such as MWC and PNS, it is the software complexity that limits the RD approach, since complexity of its recovery algorithm scales with the high Nyquist rate $W$.

The MWC provides a sampling solution for scenarios in which the input reaches frequencies that are beyond the analog bandwidths of commercial ADCs, $f_{\max} > b$. The system can be used when knowledge of the carrier positions is present or absent. Furthermore, even when $f_{\max}$ is moderate, say within T/H bandwidth $b$ of available ADC devices, the MWC proposes an advantage of reducing the processing rates, so that a cheap processor can be used instead of a premium device that can accommodate the Nyquist rate. In fact, even when the sole purpose of the system is to store the samples, the cost of storage devices that are capable of handling high-speed bursts of data streams, with or without compression, may be a sufficient motivation to shift from Fig. 1 towards the MWC sub-Nyquist system. The hardware prototype [47] is also applicable for sampling FRI signals at their innovation rate [66]. A recent publication [51] introduces Xampling, a generic framework for signal acquisition and processing in UoS. The Xampling paradigm is built upon the various insights and example applications we surveyed in this review and the general sampling approach developed in [39].

Finally, we would like to point out the Nyquist-folding receiver of [80] as an alternative sub-Nyquist paradigm. This method proposes an interesting route to sub-Nyquist sampling. It introduces a deliberate jitter in an undersampling grid, which induces a phase modulation at baseband such that the modulation magnitude depends on the unknown carrier position. We did not elaborate on [80] since a reconstruction algorithm was not published yet. In addition, the jittered sampling results in a nonlinear operator and thus departs from the linear framework of generalized sampling which unifies all the works we surveyed herein. However, this is an interesting venue for developing sub-Nyquist strategies which exploit nonlinear effects.

In an eye towards the future, we would like to draw the role sub-Nyquist systems may play in the next generation of communication systems. The traditional zero-IF and low-IF receivers are based on demodulation by a given carrier frequency $f_c$ prior to sampling. Knowledge of the carrier frequency is utilized to improve circuit properties of the receiver for the given $f_c$, at the expense of degraded performance in spectrum zones that are far from the specified frequency. For example: the oscillator that generates $f_c$ in the I/Q-demodulator can be chosen to have a narrow tuning range so as to improve the frequency stability. An active mixer whose linear range is tailored to $f_c$ is another possible design choice once the carrier is known.

In the last decade, the trend is to construct generic hardware platforms in order to reduce the production expenses involved in specifying the design for a given carrier. Two strategies that are recently being pushed forward are:

- software-defined radio (SDR) [81], where the receiver contains a versatile wideband hardware platform. The firmware is programmed to a specific $f_c$ after manufacturing, enabling the SDR to function in different countries or by several cellular operators; and
- cognitive radio (CR) [82], which adds another layer of programming, by permitting the software to adjust the working frequency $f_c$ according to high-level cognitive decisions, such as cost of transmission, availability of frequency channels, etc.

The interest in CR devices stems from an acute shortage in additional frequency regions for licensing, due to past allocation policies of spectral resources. Fortunately, studies have shown that those licensed regions are not occupied most of the time. The prime goal of a CR device is to identify these unused frequency regions and utilize them while their primary user is idle. Nowadays, most civilian applications assume knowledge of carrier frequencies so that standard demodulation is possible. In contrast, CR is an application where by definition spectral support varies and is unknown a-priori. We therefore foresee sub-Nyquist sampling playing an important role in future CR platforms. The MWC hardware, for instance, does not assume the carrier positions and is therefore designed in a generic way to cover a wideband range of frequencies. The ability to recover the frequency support from lowrate sampling may be the key to efficient spectrum sensing in CR [83].

---

**Box 5. Numerical Simulations of Sub-Nyquist Systems**

In this paper we have focused on bridging theory and practice, namely on a high-level survey and comparison of sub-Nyquist methods. Such a high-level evaluation reveals the potential performance and inherent limitations in a device-independent setting. Numerical simulations are often used for these evaluation purposes. This box highlights delicate points concerning simulation of sub-Nyquist sampling strategies.

**Hardware modeling.** A first step to numerical evaluation of an analog prototype is properly modeling the hardware components in a discrete computerized setup. For example, an analog filter can be represented by a digital filter with appropriate translation of absolute to angular frequencies. Modeling of an ADC device is a bit more tricky. In classic PNS works [22], [23], [43], [45], the ADC is modeled as an ideal pointwise sampler. However, as explained in Box 1, a commercial ADC has an analog bandwidth limitation which dictates the maximal frequency $b$ that the internal T/H circuitry can handle. In order to express the T/H limitations of the hardware, a lowpass filter preceding the pointwise sampling should be added [20]. The figure below depicts the spectrum of a single branch in a PNS setup. When discarding the analog bandwidth $b$, contents from high frequencies alias to baseband. Unfortunately, this result is misleading, since, in practice, the T/H bandwidth would eliminate the desired aliasing effect. This behavior is immediately noticed when inserting a lowpass filter with cutoff $b$ before the ideal sampler.

**Point density.** Once the hardware is properly modeled, the simulation computes samples on a grid of time points. The step size (in time) controls the accuracy of the computed samples compared with those that would have been obtained by the hardware. Clearly, if all the hardware nodes are bandlimited, then the computations can be performed at the Nyquist rate. The ADC is then visualized as a decimator at the end of the path. This option cannot be used, however, when the hardware nodes are not bandlimited. For example, in the MWC strategy, the periodic waveforms $p_i(t)$ are not necessarily bandlimited, and neither is the product $x(t)p_i(t)$, which, theoretically, consists of infinitely many shifts of the spectrum of $x(t)$. As a result the subsequent analog filtering, which involves continuous convolution between $h(t)$ and the nonbandlimited signal $x(t)p_i(t)$, becomes difficult to approximate numerically. In the simulations of [20], a simulation grid with density ten times higher than the Nyquist rate of $x(t)$ was used to estimate the MWC samples in a precision approaching that of the hardware. The figure below exemplifies the importance of correct density choice for simulation. The Fourier-series coefficients $c_{il}$ of a sign alternating $p_i(t)$ were computed over a grid that contains $r$ samples per each sign interval. Evidently, as $r$ increases the simulation density improves and the coefficients converge to their true theoretical values.

**Sensitivity check.** Hardware circuits are prone to design imperfections. Therefore, besides simulation at the nominal working point, it is important to check the system behavior at nearby conditions; recall the wishlist of Table 1. The rightmost pane demonstrates the consequence of applying the RD on a harmonic sparse input, whose tones spacing $\Delta$ does not match exactly the spacing that the system was designed for. The reconstruction error is large; see also [51], [84]. Numerical simulations in [20], [50] and hardware experiments in [47] affirm robustness of the MWC system to various noise and imperfection sources.

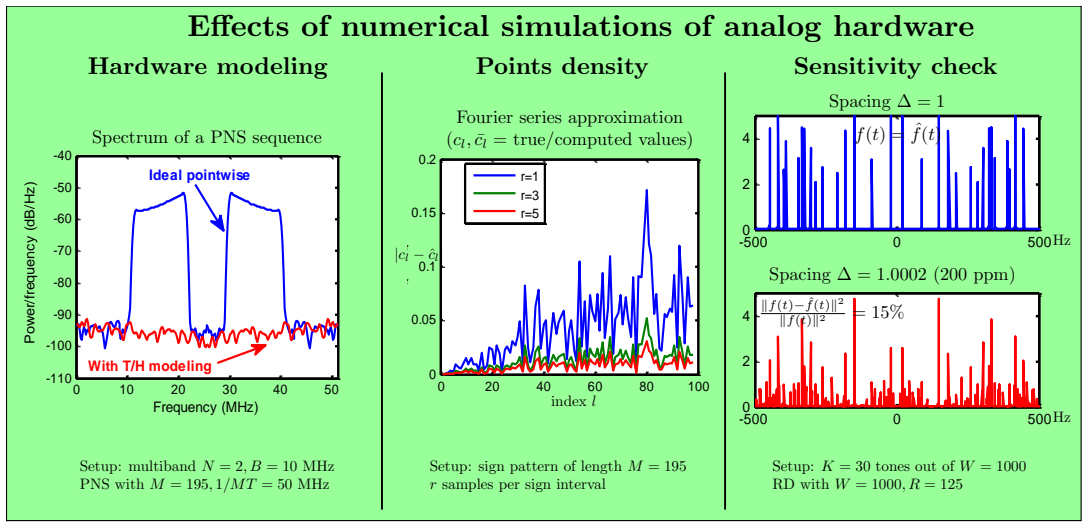

We note that discretization techniques that are used to conduct a numerical study are not to be confused with model discretization approaches which use finite signal models to begin with. As an evaluation tool, discretization gives the ability to simulate the hardware performance to desired precision. Obviously, increasing the simulation density improves accuracy, at the expense of additional computations and memory and time resources. In practice, the hardware performs analog operations instantly, regardless of the run time and computational loads that were required for numerical simulations. In contrast, when basing the approach on discretization of the analog model, the choice of grid density brings forth issues of accuracy and various complexities to the actual sampling system. Eventually, model discretization also affects the size of problems that can be simulated numerically; multiband with 10 GHz Nyquist-rate in [20] vs. a bandwidth of 32 kHz in [41].

ACKNOWLEDGMENT

The authors would like to thank the anonymous reviewers for their constructive comments and insightful suggestions which helped improve the manuscript.